\documentclass[twocolumn]{aastex701}
\usepackage{amsmath,amsfonts,amssymb}
\usepackage{graphicx}
\usepackage{setspace}
\usepackage{tocloft}
\usepackage{lineno}
\usepackage{booktabs}
\usepackage{makecell}

\newcommand{\reftab}[1]{Table~\ref{#1}}
\newcommand{\reffig}[1]{Fig.~\ref{#1}}
\newcommand{\refeqn}[1]{Eqn.~\ref{#1}}
\usepackage{tabularx}
\newcolumntype{L}[1]{>{\raggedright\let\newline\\\arraybackslash\hspace{0pt}}m{#1}} 
\newcolumntype{C}[1]{>{\centering\let\newline\\\arraybackslash\hspace{0pt}}m{#1}} 

\def\logt{\log_{10}}

\begin{document}

\title{Identifying Surface Degeneracies in Single-Visit Reflected Light Observations of Modern Earth using the Habitable Worlds Observatory}

\author[0000-0002-6387-7729]{Aiden S. Zelakiewicz}
\altaffiliation{New York Space Grant Fellow}
\affiliation{Department of Astronomy, Cornell University, 122 Sciences Drive, Ithaca, NY 14853, USA}
\affiliation{Carl Sagan Institute, Cornell University, 311 Space Sciences Building, Ithaca, NY 14853, USA}
\email{asz39@cornell.edu}

\author[0000-0003-0814-7923]{Elijah Mullens}
\affiliation{Department of Astronomy, Cornell University, 122 Sciences Drive, Ithaca, NY 14853, USA}
\affiliation{Carl Sagan Institute, Cornell University, 311 Space Sciences Building, Ithaca, NY 14853, USA} 
\email{eem85@cornell.edu}

\author[0000-0002-0436-1802]{Lisa Kaltenegger}
\affiliation{Department of Astronomy, Cornell University, 122 Sciences Drive, Ithaca, NY 14853, USA}
\affiliation{Carl Sagan Institute, Cornell University, 311 Space Sciences Building, Ithaca, NY 14853, USA}
\email{lkaltenegger@astro.cornell.edu}

\author[0000-0002-8711-7206]{Dmitry Savransky}
\affiliation{Department of Astronomy, Cornell University, 122 Sciences Drive, Ithaca, NY 14853, USA}
\affiliation{Carl Sagan Institute, Cornell University, 311 Space Sciences Building, Ithaca, NY 14853, USA} 
\affiliation{Sibley School of Mechanical and Aerospace Engineering, Cornell University, Ithaca, NY 14853, USA}
\email{ds264@cornell.edu}

\begin{abstract}
Characterizing the surface and atmosphere of Earth-like planets in reflected light is a key goal for upcoming direct imaging surveys.
NASA’s next flagship-class astrophysics mission concept, the Habitable Worlds Observatory (HWO), is a space-based Ultraviolet/Optical/Near-Infrared observatory with a mission design requirement to reach the $10^{-10}$ contrast necessary to characterize Earth-like planets around Sun-like stars.
While reflected light from planetary surfaces provides a unique opportunity to constrain the coverage of surface materials and biopigments, detailed predictions of HWO's ability to retrieve surface fractions are necessary but have not been conducted.
Here, we model photon-counting noise from astrophysical, instrumental, and post-processing sources for the HWO Exploratory Analytic Case 5 design equipped with a charge-6 vector-vortex coronagraph.
By combining our photon-counting noise with five distinct modern Earth models at quadrature, we simulate single-visit HWO observations and perform spectral retrievals using the open-source code $\texttt{POSEIDON}$ to assess our ability to constrain both the surface and atmospheric composition.
We find that degeneracies between planetary radius, surface pressure, surface material, and cloud coverage in reflected-light retrievals can significantly complicate the classification of surface features.
These degeneracies can complicate the detection of surface biopigments, such as the chlorophyll-induced red edge on modern Earth.
Our work shows that developing concrete strategies for detecting surface features and breaking degeneracies in reflected-light observations of Earth-like planets is a critical priority for mission design and data analysis.
\end{abstract}


\section{Introduction}
\label{sec:intro}  



With over 6,000 exoplanets confirmed to date~\citep{christiansen_nasa_2025}, and tens of thousands more predicted to be found with upcoming observatories like the \textit{Nancy Grace Roman Space Telescope}~\citep{wilson2023} and \textit{PLATO}~\citep{heike2025}, the field of exoplanet science has entered an era of characterization and demographic studies.
Atmospheric characterization using the \textit{James Webb Space Telescope} (\textit{JWST}) is unveiling features such as molecular abundances~\citep{bell2023,benneke2024,beatty2024,schmidt2025}, aerosols~\citep{grant2023,dyrek2024,inglis2024}, and photochemistry~\citep{tsai2023, rustamkulov2023} in transiting hot Jupiters and sub-Neptunes.
Transmission and emission spectroscopy of Earth-like planets, however, is significantly more difficult due to factors such as the shallower transit depth~\citep{kaltenegger2009}.
Despite the challenges, \textit{JWST} observations of rocky exoplanets orbiting M-dwarfs have begun to place upper limits on secondary atmospheres~\citep{xue2024,radica2025,fortune2025,bennett2025, espinoza2025, glidden2025}.
While M-dwarfs improve the observability of rocky exoplanets compared to Sun-like stars, their increased activity, especially young M-dwarfs, may expose close-in planets to atmospheric stripping due to high-energy irradiation.
Stellar activity may also present signals larger than attenuation from the planet itself, which require an in-depth understanding of the star to address~\citep{scalo2007,lammer2007,wheatley2017,kaltenegger_how_2017, zahnle2017}.

The 2020 Decadal Survey on Astronomy and Astrophysics by the National Academies of Science, Engineering, and Medicine identified imaging Earth-like planets around Sun-like stars as a top priority for the field~\citep{national_academies_of_sciences_pathways_2021}.
Following the recommendations of the 2020 Decadal Survey, NASA announced the Habitable Worlds Observatory (HWO) as the next flagship-class astrophysics mission design concept, taking inspiration from the LUVOIR~\citep{the_luvoir_team_luvoir_2019} and HabEx~\citep{gaudi_habitable_2020} concepts.
The HWO design is a $\geq6$ meter space-based observatory with a dedicated survey for the detection and characterization of Earth-like exoplanets around FGK stars~\citep{tuchow2025, peacock2025}.
A driving science goal of HWO is to characterize exoplanets around FGK stars, with particular emphasis on Earth-like planets around Sun-like stars.
To observe an Earth-twin orbiting a Sun-analog, HWO will need to reach a planet-star flux ratio ($F_p/F_s$) below $10^{-10}$ through high-contrast imaging.

As part of the Great Observatory Maturation Program (GOMAP), the Technology Assessment Group (TAG) has been leading the design of Exploratory Analytic Cases (EACs) for HWO~\citep{national_academies_of_sciences_pathways_2021, feinberg2024}.
The EACs were created to model and inform key mission design choices as the Concept Maturity Level timeline for HWO progresses.
The first three EAC design efforts, which ranged from 6 to 8 meters, were completed in 2025, with EACs 4 and 5 currently being assessed by the TAG.
The EAC 4 and 5 primary mirrors have an {inscribed} diameter of $>6.5$ and {$>8$} meters, respectively~\citep{feinberg2026}.
Both observatory designs feature hexagonal segmented primary mirrors and an off-axis secondary.
The EACs are compatible with a coronagraph instrument (CI), which will enable visible (VIS) and near-infrared (NIR) spectroscopy using an integral field spectrograph (IFS) to capture reflected light from an Earth-sized exoplanet.

In the wavelength regime of HWO, reflected light from the host star is expected to be the primary component of emergent flux from the planet.
Reflection spectroscopy of Earth-like exoplanets has the potential to uncover biosignatures unique to surface features, such as biopigments, including the chlorophyll-induced photosynthetic red edge at $\sim750$ nm~\citep{sagan1993, seager2005, kaltenegger_how_2017}.
Ongoing research on identifying potential atmospheric biosignatures on other worlds \citep[see][for reviews]{kaltenegger_how_2017, schwieterman_exoplanet_2018, schwieterman_overview_2024}, and biopigments on Earth~\citep{hegde_colors_2013, hegde_surface_2015, omalley-james_biofluorescent_2018,omalley-james_expanding_2019, coelho_color_2022, coelho_purple_2024, coelho2025} have shown the incredible spectral diversity of life.
Detecting surface biopigments on terrestrial exoplanets will require rigorous protocols for signal detection and vetting~\citep{meadows2022}.

Here, we examine whether current state-of-the-art methods for analyzing exoplanet atmospheres via {Bayesian} atmospheric retrievals \citep[see][for reviews]{madhu2018, barstow2020, macdonald_catalog_2023} can identify surface features in addition to atmospheric features.
Retrieval codes are the standard for analyzing \textit{JWST} transmission and emission spectra, and many of these algorithms are now being developed to handle higher-resolution reflection spectra for next-generation observatories.
These codes must account for the unique spectral signatures of surface pigmentation from both abiotic and biotic sources, as well as for attenuation by atmospheric gases and aerosols.

Work has begun on constructing frameworks to determine the surface composition of exoplanet reflection spectra.
The simplest and most agnostic surface treatment is to assume a constant, wavelength-independent surface albedo ~\citep{feng2018,damiano2022, salvador2025}.
Assuming a constant surface albedo reduces computational cost and model complexity, but it does not capture the unique spectral characteristics of individual surface components~\citep{madden_how_2020}.
To capture the wavelength dependence of surface reflectivity while remaining agnostic to prior information about the planet, some studies have developed schemes to analytically reproduce the planet's surface albedo, successfully placing constraints on Earth's photosynthetic red edge~\citep{wang_unveiling_2022,gomez_barrientos_search_2023, burr2026} and the overall structure of Earth's surface albedo~\citep{ulses_detecting_2025}.
However, while implementing analytical treatments of the albedo profile is ideal for capturing wavelength dependence, it can increase the dimensionality of the retrieval as more complex structures are sought to be captured, thereby substantially increasing computational time.

To determine the specific composition of the planetary surface, laboratory-measured wavelength-dependent albedo data are necessary in the retrieval process.
Retrievals on a lifeless Earth~\citep{ulses_detecting_2025} placed constraints on surface coverage using ultraviolet (UV), VIS, and NIR data.
\citet{ulses_detecting_2025} found that including UV observations down to $0.3{\rm ~\mu m}$ is necessary to break degeneracies between O$_3$ abundance, planet size, and surface properties.
However, this analysis used the same set of input spectra in the forward model as in the retrieval, and the choice of laboratory data sources can significantly bias surface spectra~\citep{roccetti_hamster_2024, roccetti_planet_2025}.
Thorough tests of the method's effectiveness using spectra from commonly used databases on unknown surface compositions are critical for assessing its validity, yet have not been done.

Here, we perform surface and atmospheric retrievals on simulated single-visit observations of a modern Earth twin to identify degeneracies in reflected-light retrievals of Earth-like exoplanets.
We chose to focus on the largest EAC design concept, a 10-meter {(8.3~meter inscribed diameter)} EAC 5, to simulate HWO observations of five distinct modern Earth spectral models.
In Section~\ref{sec:methods}, we outline our methodology for modeling the modern Earth spectra, our HWO noise model, and discuss the retrieval algorithm and initialization.
We explore the retrieval results for each of the five spectra models in Section~\ref{sec:results}.
In Section~\ref{sec:discussion}, we summarize the results in the context of current retrieval algorithms and recommendations for future work.

\section{Methods} \label{sec:methods}

To constrain the surface and atmosphere of Earth-like exoplanets, we use the open-source exoplanet retrieval code {\tt POSEIDON}~\citep{macdonald_hd_2017, macdonald_poseidon_2023}$^\text{,}$\footnote[1]{https://github.com/MartianColonist/POSEIDON} that can perform retrievals for transmission, emission~\citep{coulombe_broadband_2023}, and reflection spectroscopy~\citep{mullens2024}.
The reflection capabilities in {\tt POSEIDON} were adapted from {\tt PICASO}~\citep{batalha_exoplanet_2019} and originally described in the {\tt POSEIDON v1.2} release~\citep{mullens2024}.
Here, we use an updated forward model prescription to include wavelength-dependent surfaces using lab-derived albedos ~\citep{mullens2026}.
For a complete description of the radiative transfer and the module rework that enables HWO-specific analyses, see \citet{mullens2026}.

Modern Earth serves as an ideal testbed for this analysis, owing to the extensive work undertaken to characterize Earth and the goal of detecting Earth twins beyond the Solar System~\citep{sagan1993, kaltenegger_spectral_2007, robinson_earth_2011, rugheimer_spectral_2013, feng2018, kaltenegger_finding_2020, goodis_gordon_polarized_2024}.
We perform exoplanet retrievals on five synthetic forward-modeled modern Earth spectra, which can be categorized into two {groups}: three spectra using only {\tt POSEIDON}'s forward modeling capabilities, and two spectra based on complex atmospheric and surface models of modern Earth.
The first three spectra are generated entirely within {\tt POSEIDON} and share identical parametric descriptions of the atmospheric gas-phase, temperature, and surface properties, with the goal of varying cloud cover.
These three spectra also use the same parametric description as the retrieval algorithm, as discussed in Sec.~\ref{sec:ret}, thereby enabling testing of the retrieval algorithm's performance under varying cloud coverages.
The fourth model combines a complex atmospheric pressure-temperature (P-T) and chemical mixing profile from {\tt EXO-Prime}~\citep{kaltenegger_high-resolution_2020} with an in-depth surface prescription derived from observations with the Moderate Resolution Imaging Spectrometer (MODIS)~\citep{MODIS}.
The radiative transfer for this fourth spectrum is computed within {\tt POSEIDON}, but using the more complex atmosphere and surface models.
We source the fifth spectrum from \citet{roccetti_planet_2025}, which combines the 3D radiative-transfer code {\tt MYSTIC} ~\citep{mayer2009, mayer2010} with surface spectra from the HAMSTER~\citep{roccetti_hamster_2024} hyperspectral albedo dataset.

We define the bulk planetary properties of our terrestrial planet to have surface radius $R_p=R_\oplus$, mass $M_p=M_\oplus$, and orbital distance $r_p=1{~\rm AU}$.
The Earth twin orbits a G2V star with solar metallicity (${\rm [Fe/H]}=0$), solar $\logt g=4.4$, and effective temperature $T_{\rm eff}=5772$~K, where $g$ is the solar surface gravity in cm/s$^2$.
In accordance with mission goals for HWO~\citep{national_academies_of_sciences_pathways_2021} and current potential target stars~\citep{mamajek2024, tuchow2025, peacock2025}, we assume the system is observed at a distance of $10$ pc.

Reflected light observations for a planet are expressed as the planet-star flux ratio
\begin{eqnarray} \label{eqn:fpfs}
    \frac{F_{p,\lambda}}{F_{s,\lambda}}=A_g(\lambda)\Phi(\alpha) \left(\frac{R_p}{r_p}\right)^2,
\end{eqnarray}
where $\Phi(\alpha)$ is the planetary orbital phase function and $A_g(\lambda)$ is the geometric albedo.
The orbital phase angle $\alpha$ is defined as the angle at the planet between the star and observer (i.e., the star-planet-observer angle).
\citet{robinson_inferring_2025}, through what they called a ``happy accident,'' discovered that a normalized Henyey-Greenstein (HG) phase function~\citep{henyey1941}
\begin{eqnarray} \label{eqn:NHGphase}
    \Phi(\alpha,\bar{g})_{\rm HG} = \frac{P_{\rm HG}(\alpha,\bar{g})}{P_{\rm HG}(\alpha=0^\circ,\bar{g})},
\end{eqnarray}
with asymmetry parameter $\bar{g}=-1/3$ closely reproduces Earth's phase curve.
The HG phase function at a given phase angle and asymmetry parameter is given by
\begin{eqnarray}
    P_{\rm HG}(\alpha,\bar{g}) = \frac{1-\bar{g}^2}{\left[1+\bar{g}^2-2\bar{g}\cos(\pi - \alpha)\right]^{3/2}},
\end{eqnarray}
where the scattering angle $\Theta=\pi -\alpha$.
The optimal $\alpha$ to maximize $F_p/F_s$ occurs at the root of
\begin{eqnarray}
2\cos(\alpha)\Phi(\alpha)+\sin(\alpha) \frac{\partial\Phi(\alpha)}{\partial\alpha}=0,
\end{eqnarray}
for any choice of $\Phi(\alpha)$~\citep{brown2004}.
For the normalized HG phase function at $\bar{g}=-1/3$, we find that $\alpha_{\rm max}=60^\circ\hspace{-.35em}.\hspace{0.1em}97$ maximizes the planet-star flux ratio.
Generalizing to any asymmetry parameter in the range $\bar{g}=(-1,1)$, we find the result that the optimal phase angle for $\Phi(\alpha,\bar{g})_{\rm HG}$ can be approximately fit by a straight line with $\alpha_{\rm max} \approx 90^\circ(1+\bar{g})$.

Since the reflection spectra we produce with {\tt POSEIDON} and acquired from \citet{roccetti_planet_2025} are at full phase ($\alpha = 0^\circ$), we apply \refeqn{eqn:NHGphase} to simulate observations with HWO.
We assume the Earth-like exoplanet is observed at quadrature ($\alpha = 90^\circ$) and adopt $\bar{g}=-1/3$ from \citet{robinson_inferring_2025} to produce Earth's phase curve analytically.
When we perform retrievals on HWO datasets, we cannot assume knowledge of the planet's phase function.
Coarse constraints on $\alpha$ may be possible from orbital fitting on photometric observations prior to deep spectroscopic characterization.
During retrievals, we scale the spectra (and their associated errors) back to full phase for computational purposes only.

\subsection{{\tt POSEIDON} Consistent Spectra} \label{sec:poseidonspec}

\begin{table}[ht]
\centering
\caption{Summary of parameters used to model the three self-consistent {\tt POSEIDON} Earth-like exoplanet spectra and retrievals.}\label{tab:atmo}
\begin{tabular}{@{}lccc@{}}
\toprule
Variable                    & Unit & Input & Retrieval Prior \\  
\midrule Star & & & \\ \midrule
$T_{\rm eff}$   &   K   &   5772    & --- \\
$\logt g$ & cm/s$^2$    &   4.4 &   --- \\
${\rm[Fe/H]}$  &   dex &   0   &   --- \\
\midrule Planet & & & \\ \midrule
$R_p$                       & $R_{\oplus}$ & 1  & $\mathcal{U}(0.1,~10)$     \\
$\logt P_{\rm surf}$              & bar  & 0 & $\mathcal{U}(-3,~2)$  \\
$T$                         & K & 300 & $\mathcal{U}(100,~400)$ \\
$\alpha$    & degree  & 90 & --- \\
$r_p$ &   AU    & 1   &   ---\\
\midrule Mixing Ratios & & & \\ \midrule
$\logt{\rm H}_2{\rm O}$ &---& -3    & $\mathcal{U}(-12,~0)$ \\
$\logt{\rm O}_3$        &---& -6    & $\mathcal{U}(-12,~0)$\\
$\logt{\rm O}_2$        &---& -0.7    & $\mathcal{U}(-12,~0)$ \\
\midrule H$_2$O Clouds & & & \\ \midrule
$f_{\rm cloud}$  & --- & (0, 0.5, 1) & $\mathcal{U}(0,~1)$   \\
$\logt P_{\rm top}$  & bar & $\logt(0.5)$ & $\mathcal{U}(-4,~0)$   \\
$\Delta \logt P$  & bar & $\logt(1.2)$ & $\mathcal{U}(0.1,~3)$ \\
$\logt r_{m}$  & $\mu {\rm m}$ & $ 0$ & $\mathcal{U}(-1,~1)$ \\
$\logt {\rm H_2O}$  & --- & -3 & $\mathcal{U}(-8,~0)$ \\
\midrule Surface & & & \\ \midrule
$f_{\rm forest}$  & --- & 0.15 & $\mathcal{U}(0,~1)$   \\ 
$f_{\rm ocean}$   &---  & 0.7 &$\mathcal{U}(0,~1)$\\ 
$f_{\rm snow}$   &--- & 0.05 & $\mathcal{U}(0,~1)$  \\ 
$f_{\rm sand}$  & --- &  0.1 & $\mathcal{U}(0,~1)$  \\ 
\bottomrule
\end{tabular}
\end{table}

We first test the retrieval capabilities of {\tt POSEIDON} on HWO-quality synthetic datasets generated from three forward-model spectra created within {\tt POSEIDON}.
These forward models are constructed from the parametric descriptions available in {\tt POSEIDON}'s forward-modeling and retrieval modules, enabling an \textit{apples-to-apples} comparison of HWO synthetic data with varying cloud coverage.
We create a parametric $T=300 {\rm~K}$ isothermal atmosphere with H$_2$O, O$_2$, and N$_2$ opacity sources from HITRAN~\citep{gordon2017, gordon_hitran2020_2022} in addition to O$_3$ from \citet{serdyuchenko2014} via {\tt POSEIDON}'s temperate opacity database~\citep{lin2021}.
We assign uniform vertical mixing ratios of $\logt({\rm H}_2{\rm O})=-3$, $\logt({\rm O}_3)=-6$, and $\logt({\rm O}_2)=-0.7$, where the remainder of the atmosphere is filled with N$_2$.
N$_2$-N$_2$, N$_2$-H$_2$O, O$_2$-O$_2$, and O$_2$-N$_2$ collision-induced absorption (CIA) opacity sources from HITRAN~\citep{karman2019} are also included.
The atmosphere is constructed using a 200-layer pressure grid ranging from $10^2$ to $10^{-6}$ bar with logarithmic spacing, with the opaque reflecting surface set at 1 bar.
Unlike many pressure grids used in spectral models, which assume a constant logarithmic scale, we adopt a custom pressure grid.
Between $10^{-6}$ and $10^{-2}$ bar, we place 50 pressure layers and 150 pressure layers between $10^{-2}$ and $10^2$ bar.
The higher resolution of the pressure grid from $10^{-2}$ to $10^2$ bar allows us to resolve detailed structures in the troposphere and lower stratosphere, such as clouds, which can be thin in pressure space.
We prescribe the opaque surface as 70\% ocean, 15\% forest, 5\% snow and ice, and 10\% sand, represented by the Ocean, Aspen, Antarctic, and Sand spectra listed in \reftab{tab:surf}.
In panel (B) of \reffig{fig:ptx}, the red curve in the top panel is the weighted surface spectra used in the forward model, and the individual components are shown in the bottom panel.

\begin{table*}[]
\centering
\caption{Surface spectra used in this study, as well as their percent surface coverage in the MODIS-derived surface spectra.}\label{tab:surf}
\noindent\begin{tabularx}{\linewidth}{@{}L{0.14\linewidth}ccL{0.28\linewidth}l@{}}
\toprule
Surface Name                   & Coverage (\%) & Reference                           & Spectrum Name/Location            & ID          \\ \midrule
Ocean                          & 71.57            & \citet{kokaly_usgs_2017}          & Seawater\_Open\_Ocean SW2 lwch BECKa AREF  & SW1         \\
Aspen                          & 0.89             & \citet{kokaly_usgs_2017}          & Aspen Aspen-1 green-top ASDFRa AREF      & Aspen-1     \\
Engelmann                      & 0.89             & \citet{kokaly_usgs_2017}          & Engelmann-Spruce ES-Needls-1 ASDFRa AREF & ES-Needls-1 \\
Magnolia Grandiflora           & 2.67             & \citet{meerdink_ecostress_2019}   & Magnolia grandiflora                     & JPL231      \\
Adenostoma Fasciculatum        & 0.81             & \citet{meerdink_ecostress_2019}   & Adenostoma fasciculatum 1                & VH217       \\
Reddish brown fine sandy loam & 1.89             & \citet{meerdink_ecostress_2019}   & Reddish brown fine sandy loam            & 87P1087     \\
Acacia Visco                   & 2.24             & \citet{meerdink_ecostress_2019}   & Acacia visco                             & JPL159      \\
Rangeland                      & 9.13             & \citet{kokaly_usgs_2017}          & Rangeland L02-069 S00\% G99\% ASDFRa AREF  & L02-069     \\
Wetland                        & 0.3              & \citet{kokaly_usgs_2017}          & Wetland YNP-WT-1 AVIRISb RTGC            & YNP-WT-1    \\
Wheat Residue                  & 2.59             & \citet{jennewein_spaceborne_2024} & Wheat, Field Scale Cut Green             & 33          \\
Asphalt                        & 0.17             & \citet{meerdink_ecostress_2019}   & Construction Asphalt                     & 0674UUUASP  \\
Antarctic                      & 2.87             & \citet{grenfell_reflection_1994}  & Table. 6                                 & ---         \\
Sand                           & 3.99             & \citet{kokaly_usgs_2017}          & Quartz GDS74 Sand Ottawa BECKc AREF      & GDS74 Sand  \\ \bottomrule
\end{tabularx}
\end{table*}

Clouds play an important role in Earth's energy balance~\citep{Hartmann1992,kaltenegger_how_2017, harrison1990seasonal, ramanathan1989, wielicki1995} and can significantly influence the observed spectra of a planet by obscuring surface features and reducing the path length of reflected light~\citep{tinetti2006,kaltenegger_spectral_2007, kaltenegger_how_2017,omalley-james_vegetation_2018,wang_unveiling_2022, gomez_barrientos_search_2023, borges_detectability_2024}.
We model the clouds as a slab of uniformly mixed H$_2$O particles that cover a fraction $f_{\rm cloud}$ of our atmosphere.
The cloud structure in our simulations is defined by a top-of-cloud pressure $\logt P_{\rm top}$, cloud thickness $\Delta \logt P$, mixing ratio $\logt^{\rm aero} {\rm H_2 O}$, and aerosol mean particle radius $r_{m}$.
The optical properties of the clouds are precomputed in {\tt POSEIDON} using Mie theory~\citep{mie1908} with precomputed aerosol properties from {\tt miepython}~\citep{wiscombe1979, miepython2024}.
We adopt Mie-scattering properties for liquid H$_2$O clouds using refractive indices~\citep{hale1973} provided in the aerosol database~\citep{wakeford2015, mullens2024} native in {\tt POSEIDON} ($\geq${\tt v1.2}).
A full description of {\tt POSEIDON}'s cloud treatment can be found in \citet{mullens2024, mullens2025}.

In our forward-modeled reflection spectra, we assume the cloud slab has a mixing ratio of $\logt^{\rm aero} {\rm H}_2{\rm O}=-3$ and that the particles have a mean radius of $r_m=1~\mu {\rm m}$.
The top of the cloud slab rests at a pressure of $P_{\rm top}=0.5$ bar and extends down to $0.6$ bar for a thickness of 0.1 bar, such that $\Delta \logt P = \logt(0.6/0.5)$ bar.
We generate spectra for three $f_{\rm cloud}$ scenarios: 0\%, 50\%, and 100\%.
{\tt POSEIDON} computes fractional cloud coverage for reflected-light models by computing two spectra, one clear and one cloudy, and then taking the average of the two.
Thus, the 50\% cloud-coverage spectrum is obtained by equally weighting the 0\% and 100\% spectra.
We note that given a planet in the habitable zone of its host star with a significant liquid-water ocean and secondary atmosphere, it is unlikely for the atmosphere and observed planet surface to be completely cloud-free, and 0\% cloud-coverage case is included exclusively for modeling purposes.

\subsection{MODIS Surface} \label{sec:MODIS}

\begin{table*}[]
\caption{\label{tab:modis}Earth fractional coverage using IGPB classification scheme and the spectra used to represent each component. Spectra were primarily sourced from USGS ~\citep{kokaly_usgs_2017} and ECOSTRESS ~\citep{meerdink_ecostress_2019}, with the exception of snow~\citep{grenfell_reflection_1994} and cropland~\citep{jennewein_spaceborne_2024}. Multiple spectra types under the ``Spectral Representation'' column indicate an equal weighting of the spectra.}
\noindent\begin{tabularx}{\linewidth}{@{}L{0.24\textwidth}ccC{0.24\textwidth}@{}}
\toprule
IGBP Classification                & (\%) & Spectral Representation                                                                           & Spectra Reference \\ \midrule
Water                              & 71.57         & Ocean                                                                                             &\citet{kokaly_usgs_2017}\\
Evergreen Needleleaf Forest        & 0.49          & Englemann                                                                                         &\citet{kokaly_usgs_2017}\\
Evergreen Broadleaf Forest         & 2.36          & Magnolia Grandiflora                                                                              &\citet{meerdink_ecostress_2019}\\
Deciduous Needleleaf Forest        & 0.09          & Engelmann                                                                                         &\citet{kokaly_usgs_2017}\\
Deciduous Broadleaf Forest         & 0.58          & Aspen                                                                                             &\citet{kokaly_usgs_2017}\\
Mixed Forest                       & 0.93          & \begin{tabular}[c]{@{}c@{}}Engelmann, Aspen, \\ Magnolia Grandiflora\end{tabular}                 &\citet{kokaly_usgs_2017, meerdink_ecostress_2019}\\
Closed Shrubland                   & 0.12          & \begin{tabular}[c]{@{}c@{}}Adenostoma fasciculatum,\\ Reddish brown fine sandly loam\end{tabular} &\citet{meerdink_ecostress_2019}\\
Open Shrubland                     & 2.57          & \begin{tabular}[c]{@{}c@{}}Adenostoma fasciculatum,\\ Reddish brown fine sandly loam\end{tabular} &\citet{meerdink_ecostress_2019}\\
Woody Savanna                      & 2.24          & Acacia Visco                                                                                      &\citet{meerdink_ecostress_2019}\\
Savanna                            & 3.14          & Rangeland                                                                                         &\citet{kokaly_usgs_2017}\\
Grassland                          & 5.99          & Rangeland                                                                                         &\citet{kokaly_usgs_2017}\\
Permanent Wetland                  & 0.30          & Wetland                                                                                           &\citet{kokaly_usgs_2017}\\
Cropland                           & 2.31          & Fresh Cut Wheat Residue                                                                           &\citet{jennewein_spaceborne_2024}\\
Urban and Built-Up                 & 0.17          & Asphaltic Concrete                                                                                &\citet{meerdink_ecostress_2019}\\
Cropland/Natural Vegetation Mosaic & 0.28          & Fresh Cut Wheat Residue                                                                           &\citet{jennewein_spaceborne_2024}\\
Snow and Ice                       & 2.87          & Antarctic                                                                                         &\citet{grenfell_reflection_1994}\\
Barren or Sparsely Vegetated       & 3.99          & Quartz Ottawa Sand                                                                                &\citet{kokaly_usgs_2017}\\ \bottomrule
\end{tabularx}
\end{table*}

We compute the surface albedo for a modern Earth using MODIS aboard NASA's \textit{Terra} and \textit{Aqua} satellites.
We use land-cover maps based on the International Geosphere-Biosphere Programme (IGBP) classification scheme from the MODIS Land Cover Climate Modeling Grid Product (v6.1; MCD12C1)~\citep{MCD12C1}.
The MCD12C1 science product provides annual global land-cover classifications at a 0.5-degree spatial resolution for 2001-2023.
Due to potential issues with the MCD12Q1 land product data after 2021\footnote[2]{NASA LDOPE Case Name: \href{https://landweb.modaps.eosdis.nasa.gov/displayissue?id=781}{HC\_MCD\_24254}}, from which MCD12C1 is derived, we aggregate annual data from 2010 through the end of 2020.
We compute a latitudinally cosine-weighted average of surface coverage fraction across the 11 years of data.

\begin{figure*}[htbp]
\begin{center}
\begin{tabular}{c}
\includegraphics[width=\linewidth]{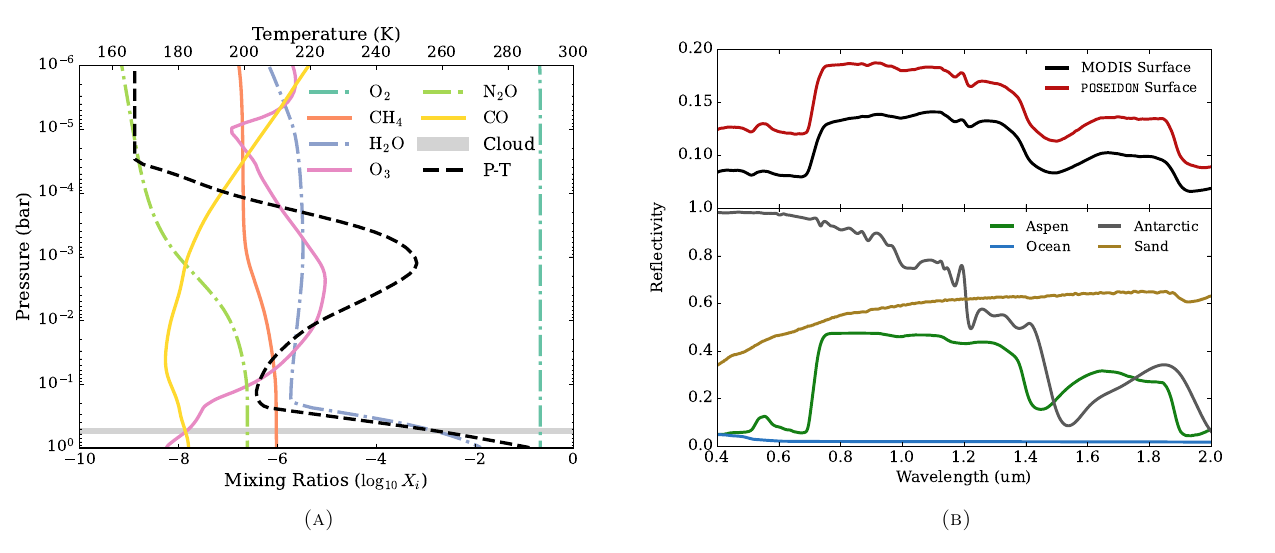}
\end{tabular}
\end{center}
\caption 
{ \label{fig:ptx} (A) P-T and chemical mixing ratio profile from \citet{kaltenegger_high-resolution_2020} used in the MODIS-derived surface coverage model. Solid and dashed-dotted curves show the vertical mixing ratios of the molecular species contributing to the spectra. The dashed black line represents the P-T profile, and the gray horizontal bar indicates the water cloud height and thickness. (B) The top panel shows the surface spectra derived from MODIS observations, as outlined in Sec.\ref{sec:MODIS}, as well as the simpler four-component surface model for the three self-consistent {\tt POSEIDON} spectra. The bottom panel shows the four spectra used in the three self-consistent forward models and in all retrievals. }  
\end{figure*}

Table~\ref{tab:modis} shows the IGPS classification scheme and surface coverage from the MODIS observations.
We assign one or more reflection spectra to represent each classification category.
The spectra we acquired are shown in \reftab{tab:surf} along with their fractional contribution to the surface reflectance spectra, shown in black in the top panel of \reffig{fig:ptx} (B).
The total contribution to the surface spectra of components that have chlorophyll-induced features (Aspen, Engelmann, Magnolia Grandiflora, Adenostoma Fasciculatum, Acacia Visco, Rangeland, Wetland, and Wheat Residue) is 19.52\%, or $f_{\rm forest}\approx0.20$.
Sand and Reddish brown fine sandy loam, which share a similar spectral structure, have a combined coverage of 6.05\%, or $f_{\rm sand}\approx0.06$.
Both of these rougher classifications are used solely to analyze the MODIS-based model's retrieval results.
Following the classification scheme in \citet{hapke_theory_2012}, the lab-derived spectra used to compute the disc-integrated reflection of exoplanets should be in the form of $r_{\rm dh}$.
However, commonly used databases, such as USGS~\citep{kokaly_usgs_2017} and ECOSTRESS~\citep{meerdink_ecostress_2019}, are constrained by laboratory data-collection methods designed for Earth and planetary science applications.
While the difference is often small, it can lead to under- or overestimation of the total reflectance (see Appendix A of \citet{mullens2026}).

The atmosphere model we used was generated using the well-tested and validated coupled 1D photochemistry and radiative transfer code {\tt EXO-Prime}~\citep{kaltenegger_spectral_2007, rugheimer_spectral_2013, rugheimer_effect_2015, kaltenegger_high-resolution_2020, madden_how_2020, madden_high-resolution_2020, kaltenegger_finding_2020}. 
We adapt the P-T and chemical mixing profile from \citet{kaltenegger_high-resolution_2020} for an Earth twin orbiting a Sun-like star, shown in panel (A) of \reffig{fig:ptx}.
While the original chemical profile contained significantly more molecules, we included only molecules for which opacity data are available in {\tt POSEIDON} and are spectrally prevalent between 0.4 and 1.8 $\mu {\rm m}$.
We include O$_2$, H$_2$O, CH$_4$, CO, and N$_2$O opacities from HITRAN~\citep{gordon2017, gordon_hitran2020_2022} and O$_3$ from \citet{serdyuchenko2014}, in addition to N$_2$-N$_2$, N$_2$-H$_2$O, O$_2$-O$_2$, and O$_2$-N$_2$ CIAs~\citep{karman2019}, using {\tt POSEIDON}'s temperate opacity database~\citep{lin2021}.
For consistency, the clouds are modeled identically to the {\tt POSEIDON} consistent case in Sec.~\ref{sec:poseidonspec} with $f_{\rm cloud}=0.5$.

\subsection{MYSTIC Model} \label{sec:mystic}

The final spectrum in our modeling set was computed by \citet{roccetti_planet_2025} using a complex, three-dimensional Earth model.
The surface prescription of this Earth model was computed using the Hyperspectral Albedo Maps dataset with high Spatial and TEmporal Resolution (HAMSTER)~\citep{roccetti_hamster_2024}.
Clouds are modeled using a 3D Cloud Generator algorithm based on liquid and ice water content data from the European Centre for Medium-Range Weather Forecast's ERA5 reanalysis product~\citep{hersbach2020}.
The spectrum was computed using the 3D Monte Carlo radiative transfer code {\tt MYSTIC}.
For more details on the creation of these spectral products, see \citet{roccetti_planet_2025}.

While \citet{roccetti_planet_2025} provides spectra at various orbital configurations, we specifically requested a spectrum at full orbital phase ($\alpha=0^\circ$) and with $\sim 50\%$ cloud coverage to remain consistent with the rest of our analysis.
For this spectrum, the 3D model's viewing geometry yielded a surface composition of 43.9\% land and 56.1\% ocean.
The land fraction is roughly 32.9\% forest, 10.1\% desert, and 0.4\% ice.
Ice and water clouds cover 41.4\% and 47.5\% of the visible planet's surface, respectively, and collectively cover 52.9\% of the visible planet.
The vertical molecular mixing ratio and P-T profile of these atmosphere models are based on the US standard atmosphere~\citep{anderson1986}.

\begin{table*}[]
\centering
\caption{\label{tab:tele}
Parameters assumed for modeling the HWO EAC 5 design. Values are adopted from the GOMAP Science-Engineering Interface repository if available and filled with values from the LUVOIR report~\citep{the_luvoir_team_luvoir_2019}.}
\begin{tabular}{@{}lllll@{}}
\toprule
Parameter                                      & Description                 & Unit                    & \multicolumn{1}{l}{VIS} & \multicolumn{1}{l}{NIR} \\ \midrule
$D_{\rm circ}$                                              & Circumscribed diameter          & m                       & \multicolumn{2}{c}{10}           \\
$D_{\rm in}$                                              & Inscribed diameter          & m                       & \multicolumn{2}{c}{8.3}    \\
$A$                                              & Collecting area          & ${\rm m}^2$                       & \multicolumn{2}{c}{65.4}    \\
$C$                                              & Coronagraph design contrast &  ---   & \multicolumn{2}{c}{$10^{-10}$}                    \\
$\mathcal{R}$                                  & Spectral resolution                  & --- & 140                     & 70              \\
$[\lambda_\mathrm{min}, \lambda_\mathrm{max}]$ & Wavelength range            & $\mu{\rm m}$            & {[}0.4,1.0{]}           & {[}1.0,1.8{]}           \\
$\theta_{\rm OWA}$                             & Outer working angle         & $\lambda/D$             & \multicolumn{2}{c}{32}                            \\
$\theta_{\rm IWA}$                             & Inner working angle         & $\lambda/D$             & \multicolumn{2}{c}{2.5}                           \\
$f_{\rm pp}$                             & Post-processing factor         & ---             & \multicolumn{2}{c}{0.1}                           \\
$T_{\mathrm{sys}}$                             & Telescope temperature       & K                       & \multicolumn{2}{c}{270}                           \\
$t_{\rm exp}$                                  & Exposure time               & hr                    & \multicolumn{2}{c}{100}                           \\
$t_{\rm read}$                                  & Read time               & s                    & \multicolumn{2}{c}{1000}                           \\
$R_\mathrm{e^-}$                               & Read noise                  & $\mathrm{ e^-}$/pixel   & 0                       & 0.4                     \\
$D_\mathrm{e^-}$                               & Dark current                & $\mathrm{ e^-}$/pixel/s & $3\times 10^{-5}$       & $1\times 10^{-4}$       \\
$R_\mathrm{C}$                                 & Clock induced charge        & $\mathrm{ e^-/pixel}$   & \multicolumn{2}{c}{$1.3\times 10^{-3}$}           \\ \bottomrule
\end{tabular}
\end{table*}

\subsection{Simulating HWO Observations} \label{sec:hwo}

\begin{figure*}[htbp]
\begin{center}
\begin{tabular}{c}
\includegraphics[width=\linewidth]{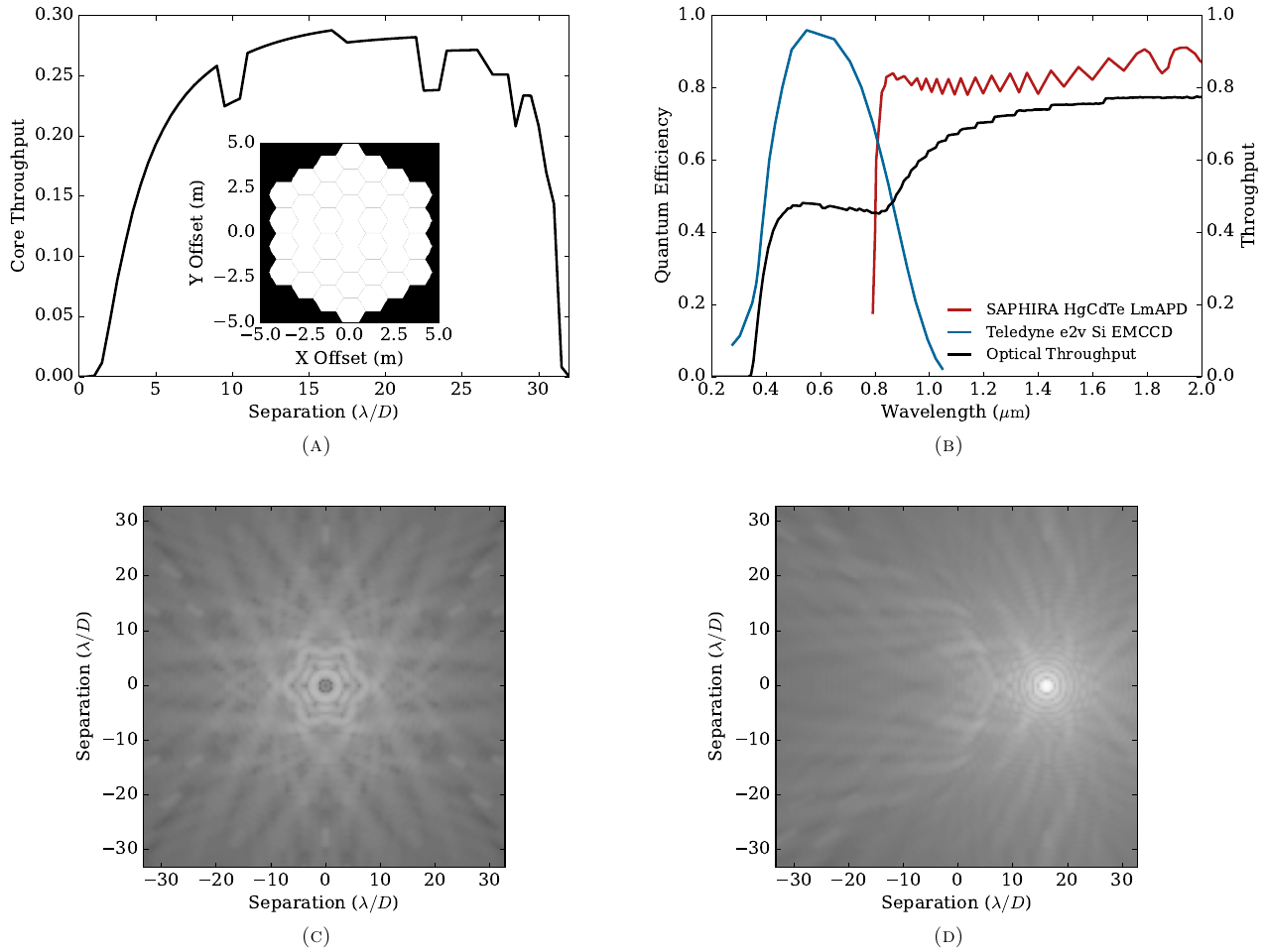}
\end{tabular}
\end{center}
\caption { \label{fig:eac5} (A) Coronagraph core throughput $\tau_{\rm core}$ and aperature of the EAC 5 design. (B) Optical throughput $\mathcal{T}$ and quantum efficiency $q$ used to model the HWO EAC 5. $\mathcal{T}$ was computed using the full EAC 1 optical path, as specified by the HWO GOMAP Science-Engineering Interface. We acquired the QE values for Teledyne e2v EMCCD (VIS) and the SAPHIRA HgCdTe LmAPD (NIR) from the GOMAP Science-Engineering Interface repository. (C) On-axis PSF of the EAC 5 aperture shown in (A). (D) Off-axis PSF of the EAC 5 aperture shown in (A) at a separation of $\lambda/D=16$.} 
\end{figure*}

The EAC 5 design features $37$ hexagonal segments with a circumscribed diameter of $D_{\rm circ}=10$ m and an inscribed diameter of $D_{\rm in}=8.3$ m.
Each hexagonal segment is $1.65$ m from point to point, arranged in three rings of mirrors surrounding the central segment, with a physical gap of $4$ mm between adjacent segments.
We model the EAC 5 aperture with {\tt HCIPy}~\citep{hcipy}, as shown in the inset of panel (A) of \reffig{fig:eac5}.
\reftab{tab:tele} lists all observatory parameters used to model the EAC 5.
The collecting area of the 37 hexagonal segments is $A=65.4{\rm~m^2}$, equivalent to an unobstructed circular pupil of diameter $9.12{\rm~m}$.
To account for reflectivities in the mirrors, we assume the EAC 5 optical path is identical to the design published in the GOMAP Science-Engineering Interface repository\footnote[3]{\href{https://github.com/HWO-GOMAP-Working-Groups/Sci-Eng-Interface}{https://github.com/HWO-GOMAP-Working-Groups/Sci-Eng-Interface}}.
We multiply the transmission and reflectivity of each optical element to acquire the total optical throughput $\mathcal{T}(\lambda)$, shown in panel (B) of \reffig{fig:eac5}.

\subsubsection{Coronagraph Instrument} \label{sec:CI}

One of the key design choices shaping the science yield of HWO will be the CI.
Here we model the HWO coronagraph as a Charge-6 Vector Vortex Coronagraph (VVC-6), which was considered for the LUVOIR~\citep{the_luvoir_team_luvoir_2019} and HabEx~\citep{gaudi_habitable_2020} mission designs.
We use the {\tt HCIPy}~\citep{hcipy} implementation of the VVC-6 and a Lyot stop that is 85\% of $D_{\rm circ}$ to model the point spread function (PSF).
Note that we do not include a custom apodizing mask or any speckle reduction from the deformable mirrors, which would reduce contrast to the $10^{-10}$ level required to image an Earth twin.
Thus, in our analysis, we set the core contrast to $C=10^{-10}$, independent of the source's angular separation.

To compute the core throughput $\tau_{\rm core}$, we propagate the PSF of the source on- and off-axis.
We compute the broadband PSF by sampling 11 wavelengths from a box bandpass centered at $0.5 {\rm\mu m}$ with a 10\% bandwidth.
We adopt the same focal ratio as the EAC 1 design, $F/\# = 20$, defined as the ratio of the focal length to diameter.
We assume the focal plane is Nyquist sampled, with a maximum separation of 32 $\lambda/D$.
An example of the on-axis PSF response of our modeled EAC 5 CI is shown in panel (C) of \reffig{fig:eac5}, with an off-axis source at 16 $\lambda/D$ shown in panel (D).

We compute $\tau_{\rm core}$ by performing aperture photometry using a circular aperture with radius $\sqrt{2}/2$ $\lambda/D$.
We utilize the exoplanet mission simulation package {\tt EXOSIMS}~\citep{savransky2016exosims, delacroix2016exosims} to process the off-axis PSF cube into the coronagraphic throughput shown in panel (A) of \reffig{fig:eac5}.
We also assume quantum efficiencies $q(\lambda)$ of the VIS and NIR IFS.
Panel (B) in \reffig{fig:eac5} shows the assumed $q(\lambda)$ for the Teledyne e2v EMCCD (VIS) and the SAPHIRA HgCdTe LmAPD (NIR), acquired from the GOMAP Science-Engineering Interface repository.

\begin{figure}[htbp]
\begin{center}
\begin{tabular}{c}
\includegraphics[width=0.95\linewidth]{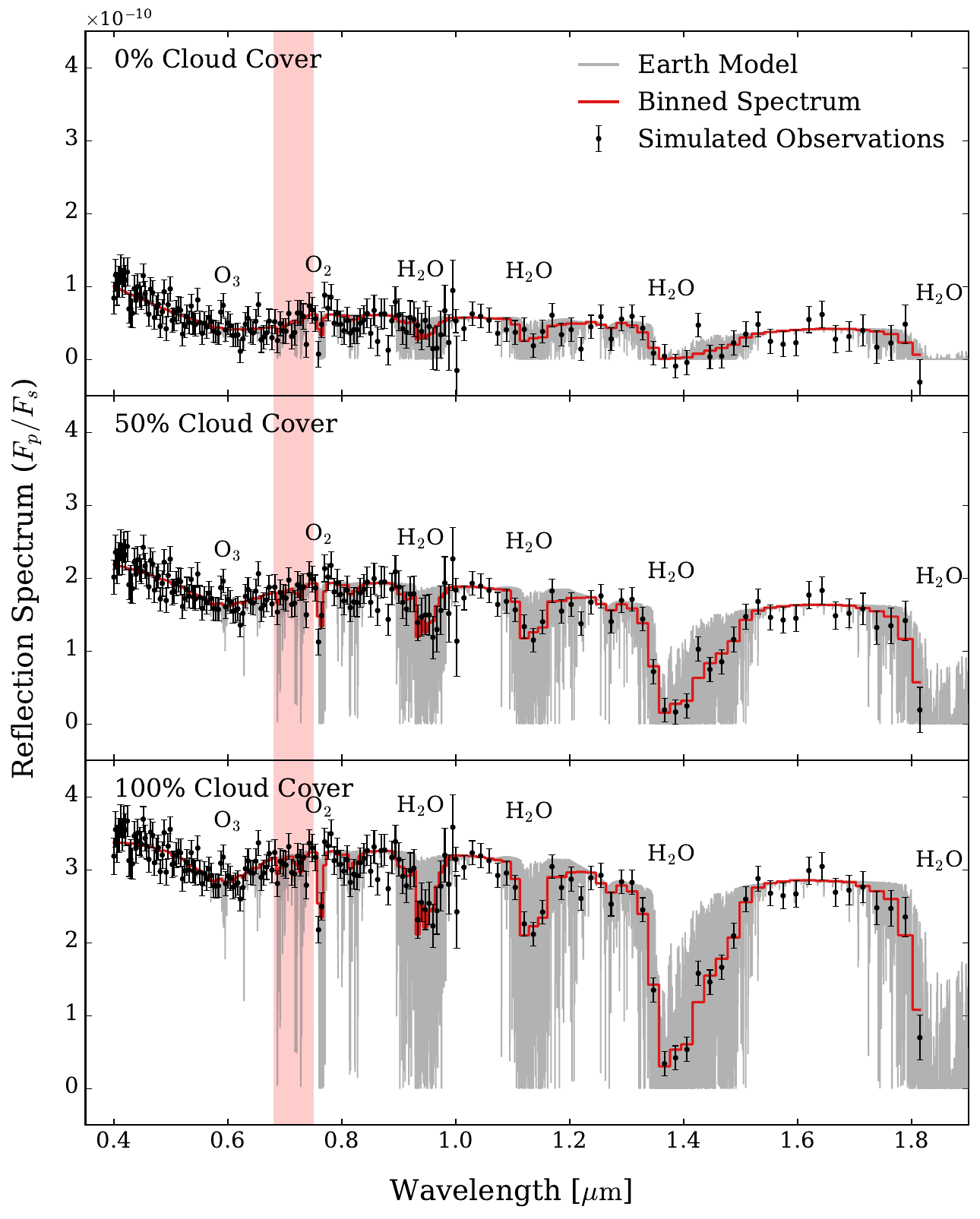}
\end{tabular}
\end{center}
\caption 
{ \label{fig:obsCloud} Simulated $t_{\rm exp}=100$ hr. observation at $\alpha=90^\circ$ of the self-consistent {\tt POSEIDON} spectra where the light gray curve denotes the high resolution ($\mathcal{R}=10^4$) forward model, the black points denote the data from the simulated HWO EAC 5 observations, and the red step curve shows the high-resolution model binned down to the simulated observation wavelength bins. Gas-phase atmospheric absorption features for H$_2$O, O$_2$, and O$_3$ are labeled along with the location of the red edge (red bar). The top, middle, and bottom panels are for the $f_{\rm cloud}=0$, $0.5$, and $1.0$ cases, respectively.} 
\end{figure}

\begin{figure}[htbp]
\begin{center}
\begin{tabular}{c}
\includegraphics[width=\linewidth]{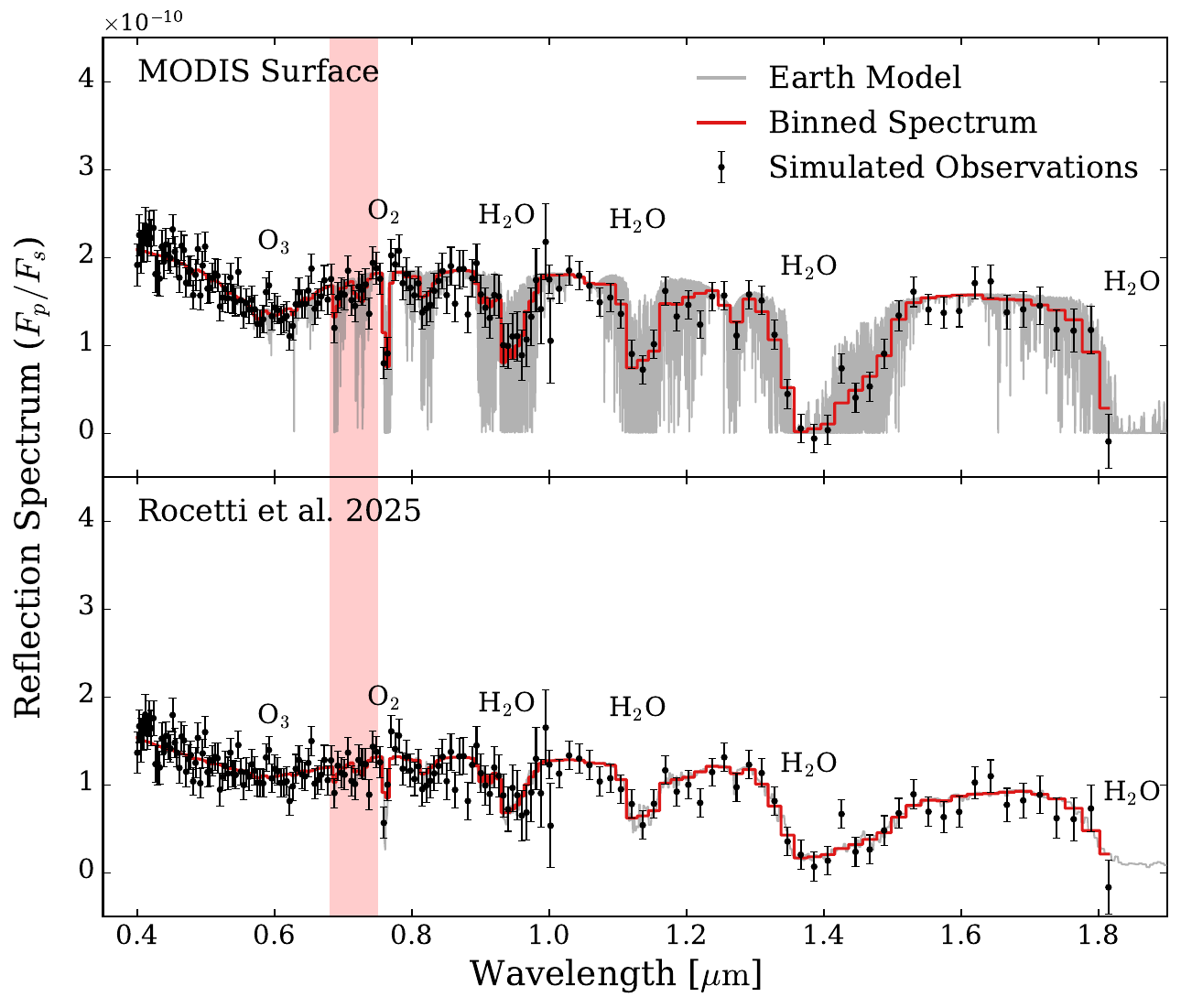}
\end{tabular}
\end{center}
\caption 
{ \label{fig:obsComplex} Simulated $t_{\rm exp}=100$ hr. observation at $\alpha=90^\circ$ of the two complex spectral models. The light gray curve shows the high resolution ($\mathcal{R}=10^4$) MODIS-derived forward-modeled spectra and the spectra from \citet{roccetti_planet_2025} in the top and bottom panels, respectively. The black points denote the data from the simulated HWO EAC 5 observations and the red step curve shows the model spectrum binned down to the simulated observation wavelength bins. Gas-phase atmospheric absorption features for H$_2$O, O$_2$, and O$_3$ are labeled along with the location of the red edge (red bar).} 
\end{figure}

\subsubsection{Photon Counting}

We estimate photon-counting errors in our simulated spectra using a modification of the noise routines included in the {\tt coronagraph} package~\citep{lustig-yaeger_coronagraph_2019}, which are based upon the implementations of \citet{brown_single-visit_2005, stark_maximizing_2014, robinson_characterizing_2016}.
For a given exposure time $t_{\rm exp}$, we compute the signal-to-noise ratio (SNR)~\citep{garrett2017} of the observation as
\begin{eqnarray} \label{eqn:snr}
    {\rm SNR} \left(t_{\rm exp}\right) = \frac{c_p}{\sqrt{(c_p+c_b)/t_{\rm exp} + c_{sp}^2}},
\end{eqnarray}
where noise in the photon count rate from the planet $c_{\rm p}$, all background sources $c_{\rm b}$, and variance in the speckle residual $c_{\rm sp}$ are considered.
The background count rate is composed of 
\begin{eqnarray} \label{eqn:c_b}
    c_{\rm b}=c_{\rm z} + c_{\rm ez} + c_{\rm sr} + c_{\rm DC} + c_{\rm RN} +c_{\rm CIC} + c_{\rm th},
\end{eqnarray}
where $c_{\rm z}$ and $c_{\rm ez}$ are count rates from zodiacal and exozodiacal dust, respectively.
The residual starlight that doesn't get suppressed by the optical system is given by $c_{\rm sr}$.
Count rates from the telescope and detector are accounted for via dark current $c_{\rm DC}$, read noise $c_{\rm RN}$, clock-induced-charge $c_{\rm CIC}$, and thermal contributions $c_{\rm th}$.
The specific forms we adopt for each count-rate term are outlined in Appendix~\ref{sec:counts}.

Single-draw Gaussian noise can be used to simulate the variation in an observation accurately.
We randomly sample a Gaussian distribution $\mathcal{N}(\mu, \sigma^2)$ to simulate a single-visit HWO observation, where the mean $\mu=(F_{p,\lambda}/F_{s,\lambda})$ is the modeled spectra and the standard deviation $\sigma=(F_{p,\lambda}/F_{s,\lambda})/{\rm SNR(t_{\rm exp})}$ is computed from our noise model.
For the HWO EAC 5 scenario we model, we assume an exposure time of $t_{\rm exp}=100$ hours.
Previous work has noted the bias that single-draw Gaussian noise can introduce into retrievals with low spectral resolution~\citep{feng2018}.
A shift in a small cluster of data points, particularly near characteristic absorption or reflection bands, can significantly affect retrieval results.
{To investigate the effects of the single-draw Gaussian noise on our retrieval algorithm, we performed an additional retrieval with the $f_{\rm cloud}=0.5$ model, fixing the noise onto the true flux values with errors set by the SNR.
We outline the results of this retrieval in Appendix~\ref{sec:fixed} and note that it ultimately did not affect the results of the retrieval.
Thus, to simulate a single observation using HWO, we proceed with the Gaussian noise draw.}

{Figs.~\ref{fig:obsCloud} and \ref{fig:obsComplex}} present our simulated 100-hour observations for the HWO EAC 5 design outlined above, for each spectral model.
The gray curves in {\reffig{fig:obsCloud} and the top panel of \reffig{fig:obsComplex}} denote the high-resolution ($\mathcal{R}=10^4$) spectra we created using the Earth models outlined in Secs.~\ref{sec:poseidonspec} and \ref{sec:MODIS}.
The gray curve in {the bottom panel of \reffig{fig:obsComplex}} shows the model from \citet{roccetti_planet_2025} with a constant wavelength bin width of $\Delta\lambda=1 {\rm~nm}$.
The red step curves show the spectra binned down to $\mathcal{R}=140$ and $\mathcal{R}=70$ in the VIS (0.4-1.0 ${\rm \mu m}$) and NIR (1.0-1.8 ${\rm \mu m}$), respectively, with the simulated observations in black points with symmetric errors given by $\sigma_\lambda(t_{\rm exp})$.

The individual noise terms used to compute the SNR, as outlined in Appendix~\ref{sec:counts}, are shown in \reffig{fig:phot} for the MODIS-derived spectra.
Individually, photons from the stellar residual $c_{\rm sr}$ and exozodiacal dust $c_{\rm ez}$ contribute significant portions of the background $c_{\rm b}$.
Due to the choice of detector technologies, read noise is absent in the VIS but contributes significantly in the NIR.
Thermal effects from the observatory are only noticeable at the reddest wavelengths, resulting in an increase in $c_{\rm b}$.
Photons from the planet, shown by the solid black curve in \reffig{fig:phot}, also consistently remain within an order of magnitude of $c_{\rm b}$.

\begin{figure}[htbp]
\begin{center}
\begin{tabular}{c}
\includegraphics[width=\linewidth]{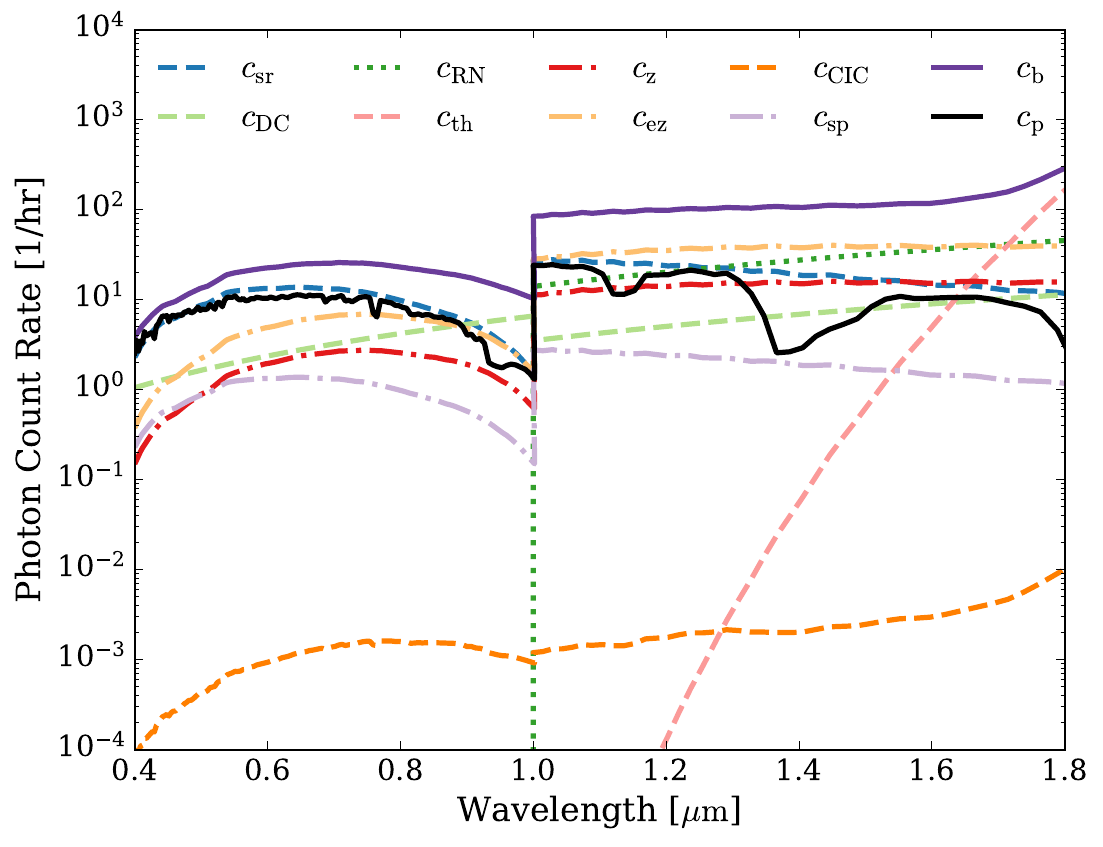}
\end{tabular}
\end{center}
\caption 
{ \label{fig:phot} Individual photon count rates for the MODIS-derived spectra. The forms of the count rates are outlined in App.~\ref{sec:counts}. The planet count rate is consistently within an order of magnitude of the total background except at the reddest wavelengths.} 
\end{figure}

\begin{figure}[htbp]
\begin{center}
\begin{tabular}{c}
\includegraphics[width=\linewidth]{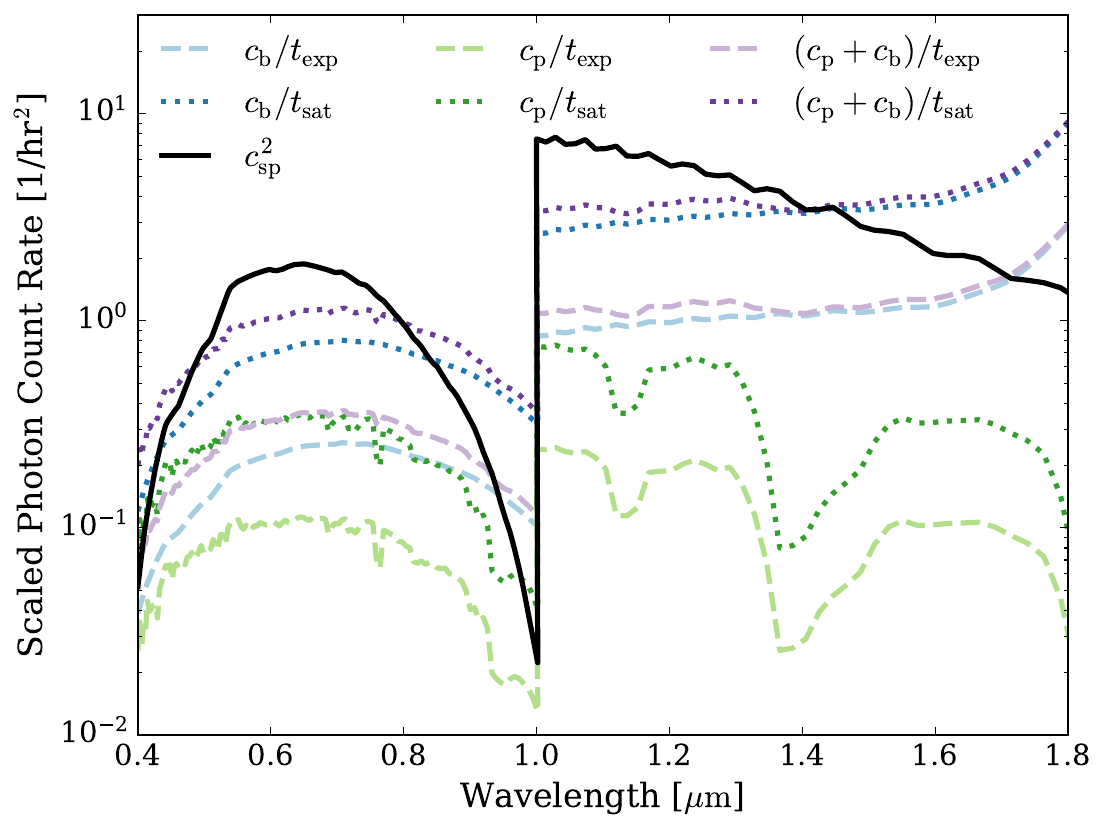}
\end{tabular}
\end{center}
\caption 
{Photon noise contributions to the simulated observations from the planet and all background sources scaled by the $t_{\rm exp}$. The solid black curve shows the square of photon counts from variations in the speckle residual, $c_{\rm sp}$, which serves as the observational noise floor. Dashed curves show the scaled count rates from the nominal exposure time of $t_{\rm exp}=100$ hr. The dotted curves show where $t_{\rm exp}$ sets the median of $c_{\rm sp}^2$ equal to the median of $(c_{\rm b} +c_{\rm p})/t_{\rm exp}$, denoted as the saturation time $t_{\rm sat}$. For our observatory configuration and $f_{\rm pp}=0.1$, the median saturation time is $\tilde{t}_{\rm sat}=32$ hrs.
\label{fig:phot_scale}} 
\end{figure}

Variations in the stellar residual $c_{\rm sp}$ dominate the noise in the observed spectra when the exposure time increases significantly, as shown in \refeqn{eqn:snrlim}.
When contributions from speckle residual variations dominate the observation noise, any increase in $t_{\rm exp}$ will not improve the SNR.
We denote this turnover as the saturation time $t_{\rm sat}$.
The saturation time is set by finding the $t_{\rm exp}$ that satisfies 
\begin{eqnarray}
    (c_{\rm p} + c_{\rm b})/t_{\rm exp} = c_{\rm sp}^2.
\end{eqnarray}
For our observatory model, target system, and expected post-processing ability, we find a median saturation time of $\tilde{t}_{\rm sat}=32 {\rm~hrs}$ for the MODIS-derived spectra.
For the observing scenario considered in our work, a 100-hr exposure time greatly reduces the noise in $c_{\rm p}$ and $c_{\rm b}$, ensuring that we are near the noise floor for most of the spectral coverage.
\reffig{fig:phot_scale} shows contributions to the SNR from $c_{\rm p}$ and $c_{\rm b}$ scaled by $t_{\rm exp}$ and the noise floor $c_{\rm sp}^2$ for the MODIS-derived spectra.
Dashed and dotted curves show the 100 hr exposure time used to simulate observations in our work and the median saturation time of 32 hrs, respectively.
Except near $1~\mu{\rm m}$ in the VIS, where the QE of the Teledyne e2v EMCCD degrades, and toward $1.8~{\rm\mu m}$, where thermal effects become significant, $c_{\rm sp}$ dominates noise in the observed spectrum.
At the median saturation time, we find that $c_{\rm sp}$ dominates the spectra except where the QE of the EMCCD, shown in panel (B) of \reffig{fig:eac5}, drops significantly in the VIS.
In the NIR, the combination of thermal effects towards $1.8~{\rm\mu m}$ and the decrease in photon counts from the stellar residual $c_{\rm sr}$ relative to the other photon sources results in the turnover at $1.4~{\rm \mu m}$.

\subsection{Retrieval} \label{sec:ret}

We use the same retrieval setup for each simulated observation.
We assume uniform mixing ratios of H$_2$O, O$_3$, and O$_2$ and an isothermal atmosphere.
{In the initial stages of this study, we utilized a more complex, two-gradient P-T profile~\citep{macdonald_trident_2022} to retrieve atmospheric structure.
We found that the two-gradient P-T profile produced an isothermal atmosphere regardless of the spectral model, motivating our decision to use an isothermal atmosphere to reduce the retrieval dimensionality.}
Clouds are defined by a uniformly mixed slab of Mie-scattering H$_2$O molecules in our simulations.
We set the fractional coverage for the four surface spectra shown in panel (B) of \reffig{fig:ptx} as free parameters in the retrieval.
We sample the atmosphere using the same custom 200-layer pressure grid as in the forward models.
This parameterization is identical to the three forward-modeled spectra we computed solely within {\tt POSEIDON}, as outlined in Sec.~\ref{sec:poseidonspec}.
During the retrieval process, we compute spectra at high spectral resolution ($\mathcal{R}=10^4$) and down-sample them to the simulated observing wavelengths.

We assume semi-informed priors for $R_p$ and $\logt P_{\rm surf}$.
The $\mathcal{U}(-3,~2)$ prior on $\logt P_{\rm surf}$ is informed by the $\sim 1~{\rm mbar}$ Martian and $\sim100~{\rm bar}$ Venusian surfaces.
The determination of $R_p$ is unfeasible for photometric observations alone.
Given multiple epochs of photometric observations at different orbital phases and assuming the orbital parameters of the planet are well constrained, $R_p$ may be constrained to within a factor of 2 of its actual value~\citep{gaudi_habitable_2020}.
This would enable tighter prior bounds on $R_p$, but ultimately rely on the determination of $\Phi(\alpha)$ and $r_p$.
Thus, we are unable to place tight priors on $R_p$, and assume a conservative, broad $\mathcal{U}(0.1,~10)$ prior on $R_p$ roughly informed by exoplanet radius distributions~\citep{fulton2017, kopparapu_exoplanet_2018, neil2020}.
{In Appendix~\ref{sec:limR}, we performed a retrieval using a tighter $\mathcal{U}(0.5,~2.0)$ prior on $R_p$ for the $f_{\rm cloud}=0.5$ model and discussed it in the context of the $\mathcal{U}(0.1,~10)$ prior.}
The radius retrieved is the reference radius used to solve for hydrostatic equilibrium, which we set to the lower surface pressure prior limit of 1 mbar, and is not the radius at $\logt P_{\rm surf}$.
To compute $F_p/F_s$, we set the radius of the reflecting surface as a wavelength-dependent photospheric radius, defined as the location where the optical depth reaches $\tau =2/3$.
Using the photospheric radius in emission spectrum retrievals has been shown to reduce biases in the location of the emitting surface~\citep{fortney2019, taylor2022}.
The 13 other parameters are assigned uniform priors with limits that cover the space of likely parameter values, listed in \reftab{tab:atmo}, for a total of 15 free parameters in the retrieval.

{\tt POSEIDON} performs Bayesian inference using the nested sampling algorithm {\tt PyMultiNest}~\citep{buchner2014,buchner2016}, which is a Python wrapper of the widely used {\tt MultiNest}~\citep{Feroz2009} software.
We ran our retrievals using 2000 live points ($N_{\rm live}$), which follows recommendations to utilize at least $N_{\rm live} \geq 1000$ from the original {\tt MultiNest} documentation~\citep{Feroz2009} and recent studies on the algorithm's performance~\citep{dittmann2024}.
Each retrieval ran in parallel across 96 cores on Intel Xeon Platinum 8468H 64-bit processors.
{For a further description of the {\tt MultiNest} setup used in this work, see \citet{macdonald_hd_2017}.}

\section{Results} \label{sec:results}
We simulate the performance of HWO in identifying the surfaces and atmospheres of terrestrial exoplanets across models of increasing complexity using the open-source spectral-retrieval code {\tt POSEIDON}.
We begin with the three spectra parameterizations produced by {\tt POSEIDON}'s forward model capabilities, which are identical to the retrieval parameterization.
We then extend to two complex models that employ complex atmospheric and surface structures.

\subsection{{\tt POSEIDON} Consistent Spectra}

\subsubsection{Cloud-free}

\begin{figure*}[htbp]
\begin{center}
\begin{tabular}{c}
\includegraphics[width=\linewidth]{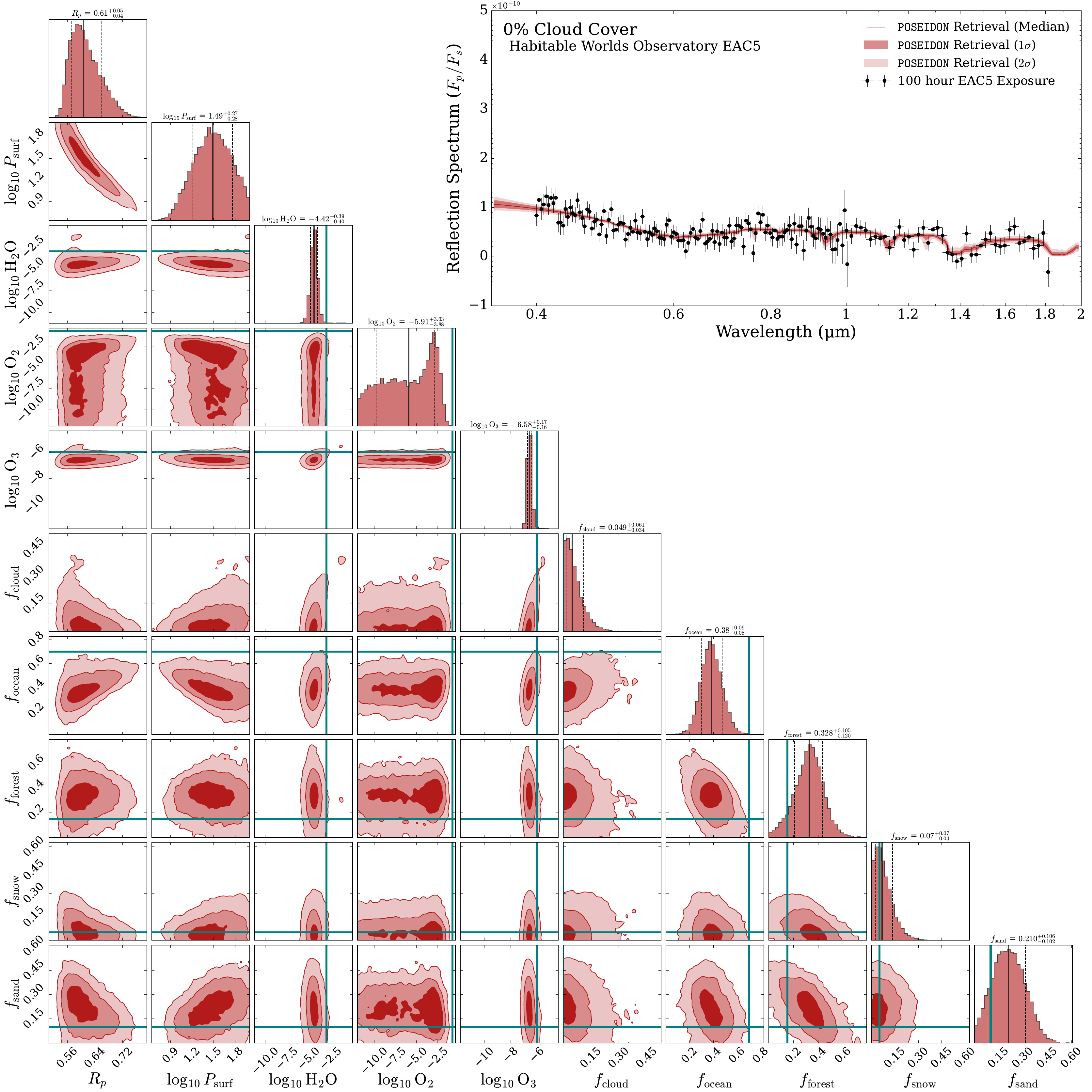}
\end{tabular}
\end{center}
\caption 
{ \label{fig:ret0cloud} Retrieval results for the self-consistent {\tt POSEIDON} spectra with 0\% cloud coverage. The posterior distributions for selected parameters are shown on the left-hand side, along with the median-retrieved spectra and simulated data in the top-right panel. The marginalized median and 1-$\sigma$ intervals for each parameter are labeled above the distributions and are shown in the black solid and dashed lines, respectively. The values used in the forward model, listed in \reftab{tab:atmo}, are shown by the solid teal line in the posteriors. The full posterior distribution is shown in \reffig{fig:corner0cloud}.} 
\end{figure*}

\reffig{fig:ret0cloud} shows the observed and retrieved spectra for the 0\% cloud coverage scenario.
The retrieval confidently computes a cloud-free planet, with $f_{\rm cloud}=0.049^{+0.061}_{-0.034}$.
In the absence of clouds, photons entering the atmosphere can reach the surface and be reflected back to the observer.
The cloud-free case, although unlikely for an Earth-like planet in the HZ with a liquid-water ocean, provides complete spectral access to the surface.
Classification of individual surface features, however, is dependent on proper constraints of $R_p$ and $\logt P_{\rm surf}$.
While the retrieval correctly identified a cloudless planet, a smaller rocky planet with $R_p=0.63^{+0.05}_{-0.03}$ and a brighter surface was favored over a larger radius with a dimmer surface.

The degeneracies identified in the retrieval are evident in the posterior distributions of $R_p$ shown in \reffig{fig:ret0cloud}.
As $R_p$ decreases and dims the spectrum by reducing the reflecting surface area, a reduction in $f_{\rm ocean}$ and an increase in  $f_{\rm sand}$ are necessary to normalize the spectrum to be equivalent to the data.
The brighter surfaces were overestimated, with $f_{\rm forest}=0.328^{+0.105}_{-0.124}$ and $f_{\rm sand}=0.210^{+0.106}_{-0.102}$, while the ocean coverage of $f_{\rm ocean}=0.38^{+0.09}_{-0.08}$ was underestimated.
The snow coverage was retrieved within 1-$\sigma$ of its input value of 0.05, with $f_{\rm snow}=0.07^{+0.07}_{-0.04}$.
The retrieval also preferred a larger surface pressure, with $\logt P_{\rm surf}=1.54^{+0.22}_{-0.22}$.
H$_2$O and O$_3$ were constrained with abundances of $\logt {\rm H_2O}=-4.42^{+0.29}_{-0.40}$ and $\logt {\rm O_3}=-6.58^{+0.17}_{-0.16}$, lower than the forward model parametrization of -3 and -6, respectively.
An upper limit on molecular O$_2$ was placed of $\logt {\rm O_2}=-5.91^{+3.03}_{-3.88}$, which is significantly lower than the input abundance of -0.7.
The inconsistency in O$_2$ is likely due to the random noise draw for the simulated observation, as one of the data points in the oxygen A band (0.76 $\mu {\rm m}$) was shifted close to the continuum level.
The planet's isothermal temperature was also significantly underestimated, and only a lower limit of $T=213^{+105}_{-74}$ K could be established, as shown in the full corner plot in \reffig{fig:corner0cloud}.

\subsubsection{Patchy Clouds} \label{sec:ret50cloud}

\begin{figure*}[htbp]
\begin{center}
\begin{tabular}{c}
\includegraphics[width=\linewidth]{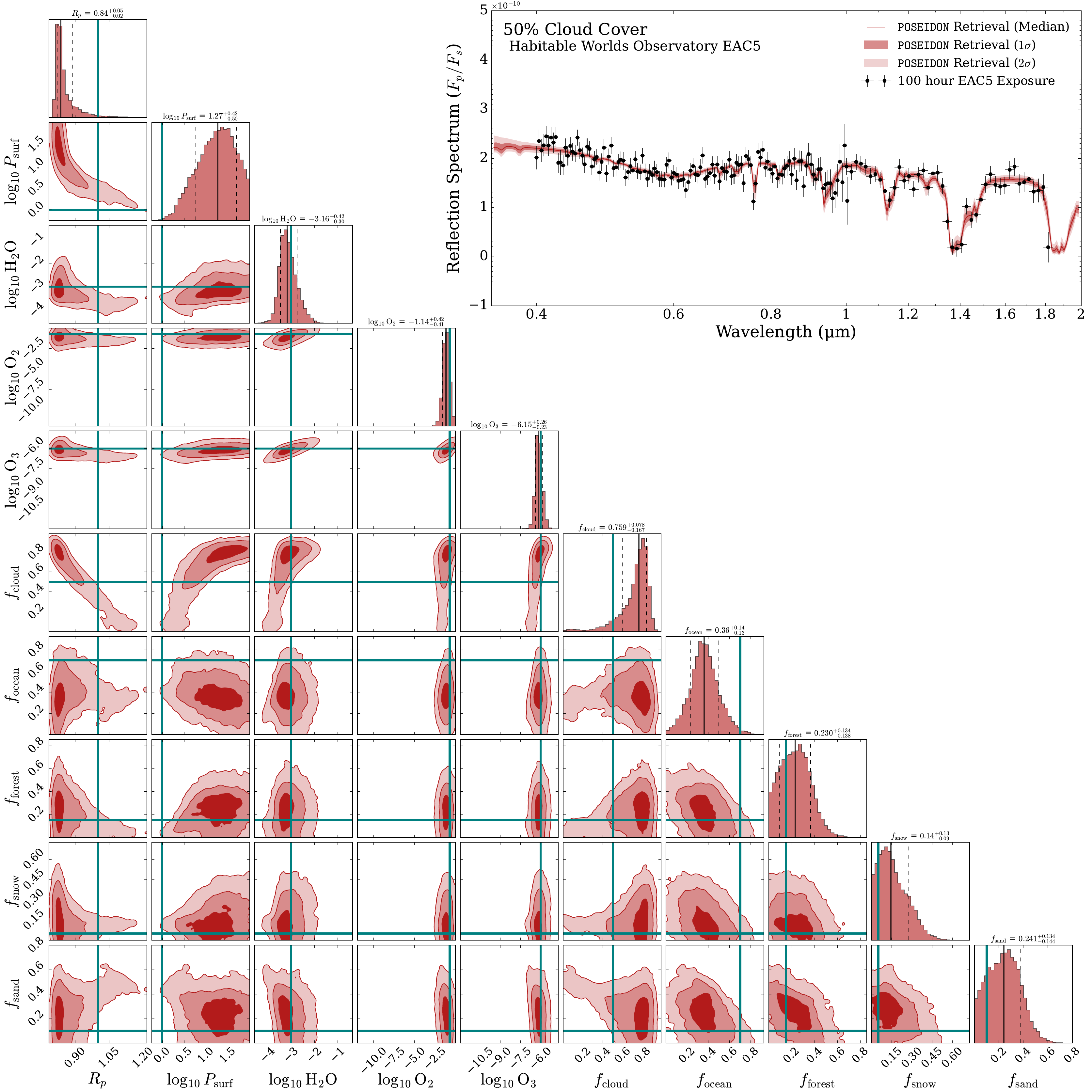}
\end{tabular}
\end{center}
\caption 
{ \label{fig:ret50cloud} Retrieval results for the self-consistent {\tt POSEIDON} spectra with 50\% cloud coverage. The posterior distributions for selected parameters are shown on the left-hand side, along with the median-retrieved spectra and simulated data in the top-right panel. The marginalized median and 1-$\sigma$ intervals for each parameter are labeled above the distributions and are shown in the black solid and dashed lines, respectively. The values used in the forward model, listed in \reftab{tab:atmo}, are shown by the solid teal line in the posteriors. The full posterior distribution is shown in \reffig{fig:corner50cloud}. } 
\end{figure*}

Retrieval results of the 50\% cloud coverage model are shown in \reffig{fig:ret50cloud}.
The retrieval model determined an upper limit of $f_{\rm cloud}=0.759^{+0.078}_{-0.167}$, indicating patchy cloud coverage.
A lower limit on the planetary radius was placed with $R_p=0.84^{+0.05}_{-0.02}$ $R_\oplus$, larger than the radius found for the 0\% cloud coverage model but still inconsistent with the 1 $R_\oplus$ for Earth.
The detection of clouds enabled the tighter constraints on $R_p$ that are closer to $1~R_\oplus$, but {a significant degeneracy between $R_p$ and $f_{\rm cloud}$ is shown clearly in the posteriors in} \reffig{fig:ret50cloud}.
The degeneracy between $R_p$ and surface coverage was now accompanied by a degeneracy in cloud coverage; clouds attenuated the surface spectra and were used by the retrieval to brighten the spectra in place of $R_p$.
This degeneracy with $f_{\rm cloud}$ is less pronounced than that associated with $R_p$, but it still affects the ability to constrain the surface accurately.
All three land spectra were constrained consistently within their input values.
We find that upper limits of $f_{\rm forest}=0.230^{+0.134}_{-0.138}$, $f_{\rm snow}=0.14^{+0.13}_{-0.09}$, and $f_{\rm sand}=0.241^{+0.134}_{-0.144}$ were placed in the retrieval, within 1-$\sigma$ of their input values of 0.15, 0.06, and 0.10, respectively.
However, the larger estimates of land fractions required a lower liquid-water ocean coverage, $f_{\rm ocean}=0.36^{+0.14}_{-0.13}$.
We find that $\logt {\rm H_2O}=-3.16^{+0.42}_{-0.30}$, $\logt {\rm O_2}=-1.14^{+0.42}_{-0.41}$, and $\logt {\rm O_3}=-6.15^{+0.26}_{-0.23}$ were all consistent with the model inputs of -3, -0.7, and -6, respectively.
The brighter continuum enhances the signal from molecular absorptions, enabling consistent constraints on gas-phase absorptions.

\subsubsection{Cloudy Day}

\begin{figure*}[htbp]
\begin{center}
\begin{tabular}{c}
\includegraphics[width=\linewidth]{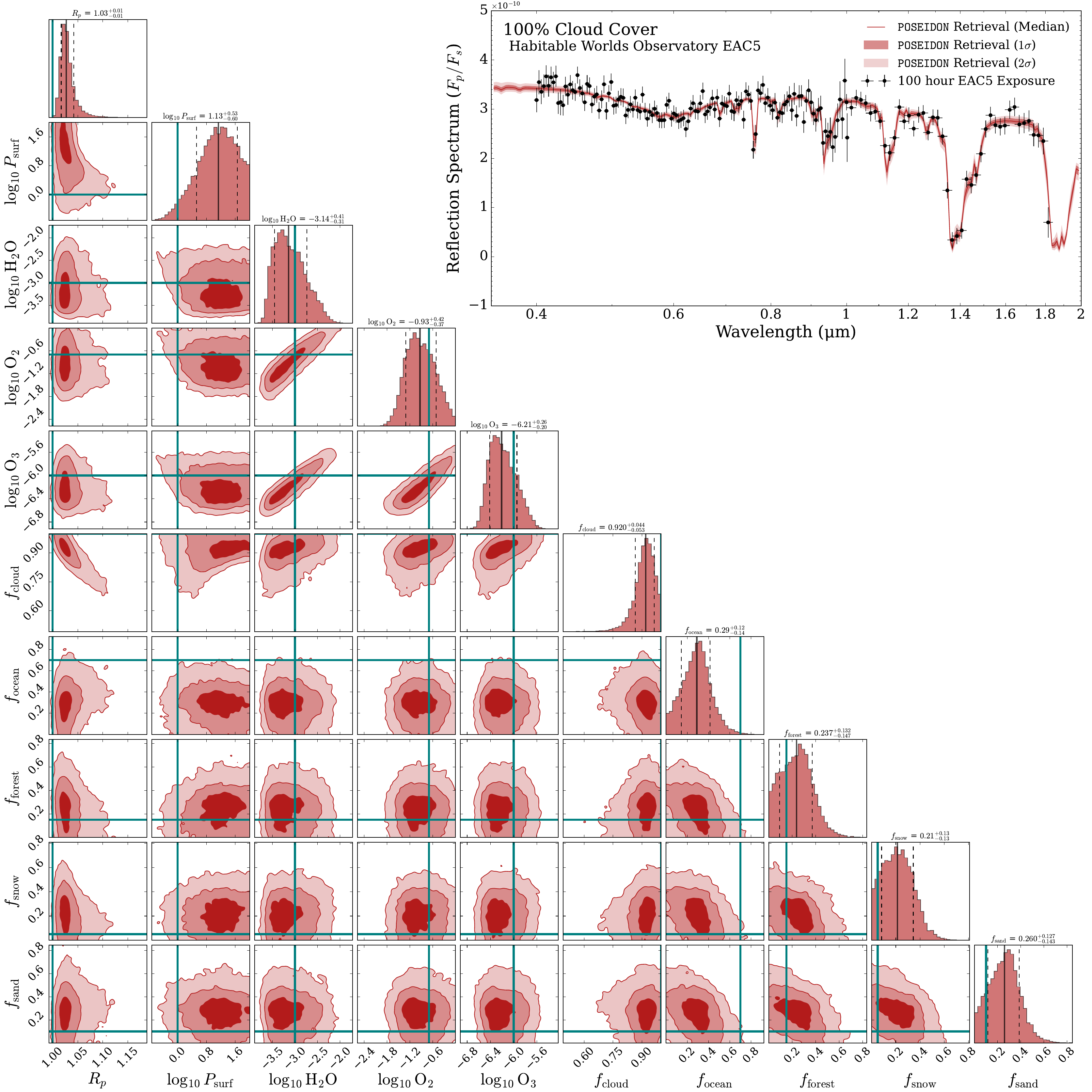}
\end{tabular}
\end{center}
\caption 
{ \label{fig:ret100cloud} Retrieval results for the self-consistent {\tt POSEIDON} spectra with 100\% cloud coverage. The posterior distributions for selected parameters are shown on the left-hand side, along with the median-retrieved spectra and simulated data in the top-right panel. The marginalized median and 1-$\sigma$ intervals for each parameter are labeled above the distributions and are shown in the black solid and dashed lines, respectively. The values used in the forward model, listed in \reftab{tab:atmo}, are shown by the solid teal line in the posteriors. The full posterior distribution is shown in \reffig{fig:corner100cloud}.} 
\end{figure*}

The retrieval results for the 100\% cloud coverage model are shown in \reffig{fig:ret100cloud}.
We find significant cloud coverage in the simulated observation, with $f_{\rm cloud}=0.92^{+0.05}_{-0.20}$ of the atmosphere being covered.
The increased continuum from the water clouds enabled $R_p=1.03^{+0.01}_{-0.01}$ to also be constrained consistently with the $1~R_\oplus$ input.
We find that the surface pressure is not as tightly constrained as $R_p$, with $\logt P_{\rm surf}=1.13^{+0.53}_{-0.60}$.
The significant cloud coverage prevents any photons from reaching the surface and reflecting to the observer.
Thus, apparent constraints on the surface are not actually informed by the spectrum, and any apparent consistency with the inputs is coincidental.
This can be seen clearly with $f_{\rm snow}=0.21^{+0.13}_{-0.13}$, which has very distinctive spectral features in the VIS and NIR, and was constrained correctly for both the 0\% and 50\% cloud coverage forward models.
The gas-phase vertical mixing ratios $\logt {\rm H_2O}=-3.14^{+0.41}_{-0.31}$, $\logt {\rm O_2}=-0.93^{+0.42}_{-0.37}$, and $\logt {\rm O_3}=-6.21^{+0.26}_{-0.20}$ were all retrieved within 1-$\sigma$ of the input values of -3, -0.7, and -6, respectively.
Constant vertical mixing ratios across the entire pressure grid ensure that features stand out above the bright cloud deck, enabling strong detection.

\subsection{External Models}

Two retrievals were run on models in which the parameters used in the forward model do not perfectly match those in the retrieval.
Thus, for the external models, we are looking to constrain specific patterns.
For the atmosphere, we want to assess the detection of 50\% cloud coverage in both models and atmospheric mixing ratios consistent with modern Earth conditions.
For MODIS-derived spectra, and even more so for the MYSTIC spectra, we are not able to reproduce the input surface albedo perfectly.
{We aim to identify two characteristics: a significant liquid-water ocean and partial forest coverage.}
Constraining the planetary radius will also be necessary to place the cloud and surface properties in context.

\subsubsection{MODIS Surface}

\begin{figure*}[htbp]
\begin{center}
\begin{tabular}{c}
\includegraphics[width=\linewidth]{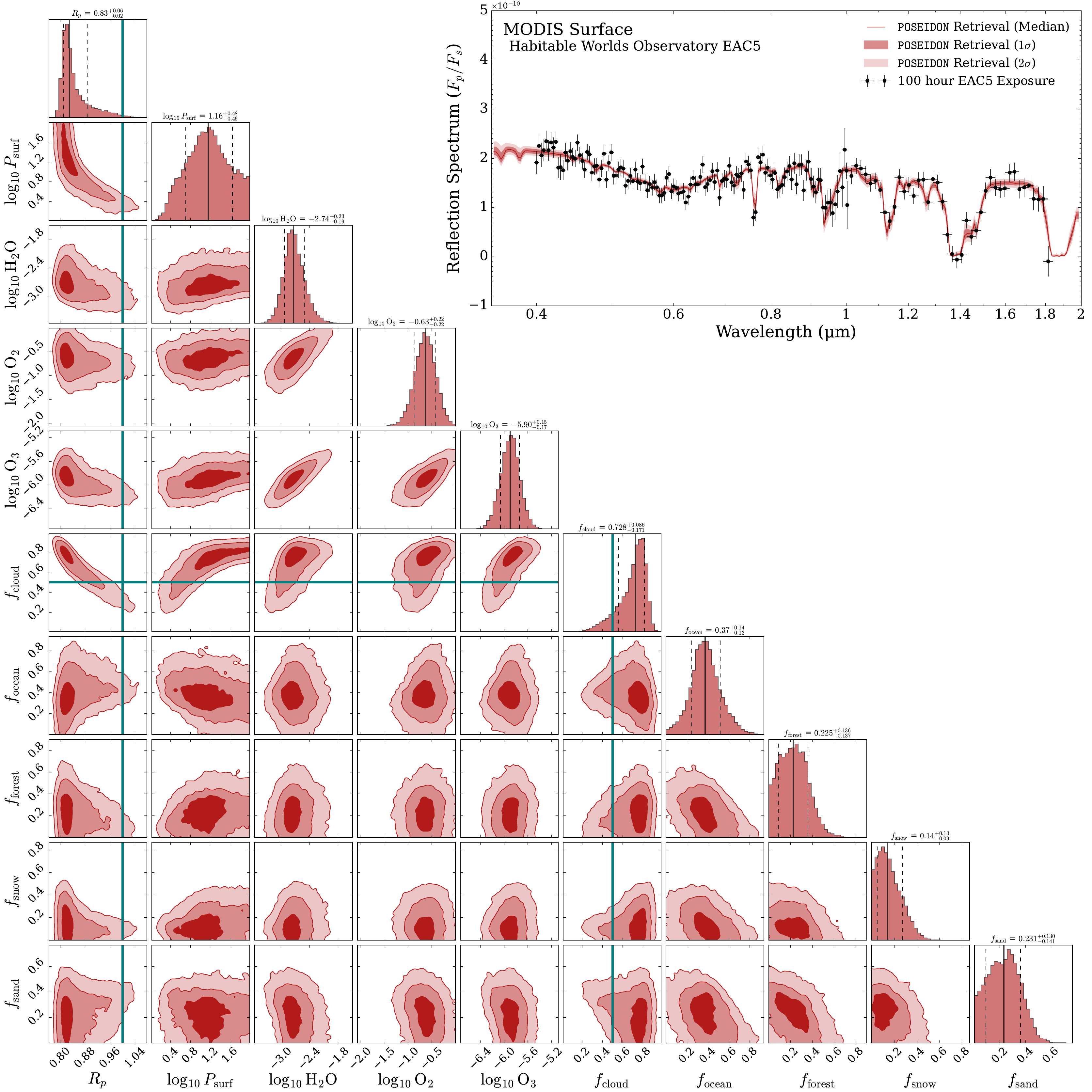}
\end{tabular}
\end{center}
\caption 
{ \label{fig:retmodis} Retrieval results for the MODIS-derived spectral model. The posterior distributions for selected parameters are shown on the left-hand side, along with the median-retrieved spectra and simulated data in the top-right panel. The marginalized median and 1-$\sigma$ intervals for each parameter are labeled above the distributions and are shown in the black solid and dashed lines, respectively. The values used in the forward model, where available, are shown by the solid teal line in the posteriors. The full posterior distribution is shown in \reffig{fig:cornerMODIS}.} 
\end{figure*}

The retrieval of simulated observations of the combined MODIS-derived surface spectra and \citet{kaltenegger_high-resolution_2020} atmospheric profile Earth model is shown in \reffig{fig:retmodis}.
An upper limit on a patchy cloud slab was detected with a coverage of $f_{\rm cloud}=0.728^{+0.086}_{-0.171}$.
We find a lower limit on the planetary radius of $R_p=0.83^{+0.06}_{-0.02}~R_\oplus$ and an overestimation of the surface pressure $\logt P_{\rm surf}=1.16^{+0.48}_{-0.46}$ bar due to the increase in cloud coverage.
\reffig{fig:retmodis} clearly shows the degeneracy between $R_p$, $\logt P_{\rm surf}$, and $f_{\rm cloud}$; if the cloud fraction were to have been at the 50\% level consistent with the input model, the retrieval would've required a larger $R_p$ and lower $\logt P_{\rm surf}$.
We find that constant vertical mixing ratios used in the retrieval model accurately constrain the complex chemical mixing profile from \citet{kaltenegger_high-resolution_2020}, shown in \reffig{fig:ptx}.
We detect $\logt {\rm H_2 O}=-2.74^{+0.23}_{-0.19}$, consistent with the mixing ratio just above the cloud deck.
Atmospheric mixing ratios for $\logt {\rm O_2}=-0.63^{+0.22}_{-0.22}$ and $\logt {\rm O_3}=-5.90^{+0.15}_{-0.17}$ were also consistent with the chemical profile used.

For this retrieval, ocean coverage was underestimated with $f_{\rm ocean}=0.37^{+0.14}_{-0.13}$ and sand surface coverage was overestimated with an upper limit of $f_{\rm sand}=0.231^{+0.130}_{-0.141}$ compared to their expected values of $0.716$ and $\sim0.06$, respectively.
As seen in \reffig{fig:ptx}, both the quartz sand and ocean spectra are relatively featureless between 0.4 and 1.8 $\mu{\rm m}$.
The increase in sand coverage and decrease in ocean coverage brightened the spectra, compensating for the smaller retrieved value of $R_p$.
Chlorophyll features were accurately detected, with $f_{\rm forest}=0.225^{+0.136}_{-0.137}$ consistent with the expected coverage of $\sim0.20$.
Snow coverage, while not consistent with the input value of $0.03$, was disfavored by the retrieval and constrained to $f_{\rm snow}=0.14^{+0.13}_{-0.09}$.

\subsubsection{MYSTIC Model} \label{sec:retmystic}

\begin{figure*}[htbp]
\begin{center}
\begin{tabular}{c}
\includegraphics[width=\linewidth]{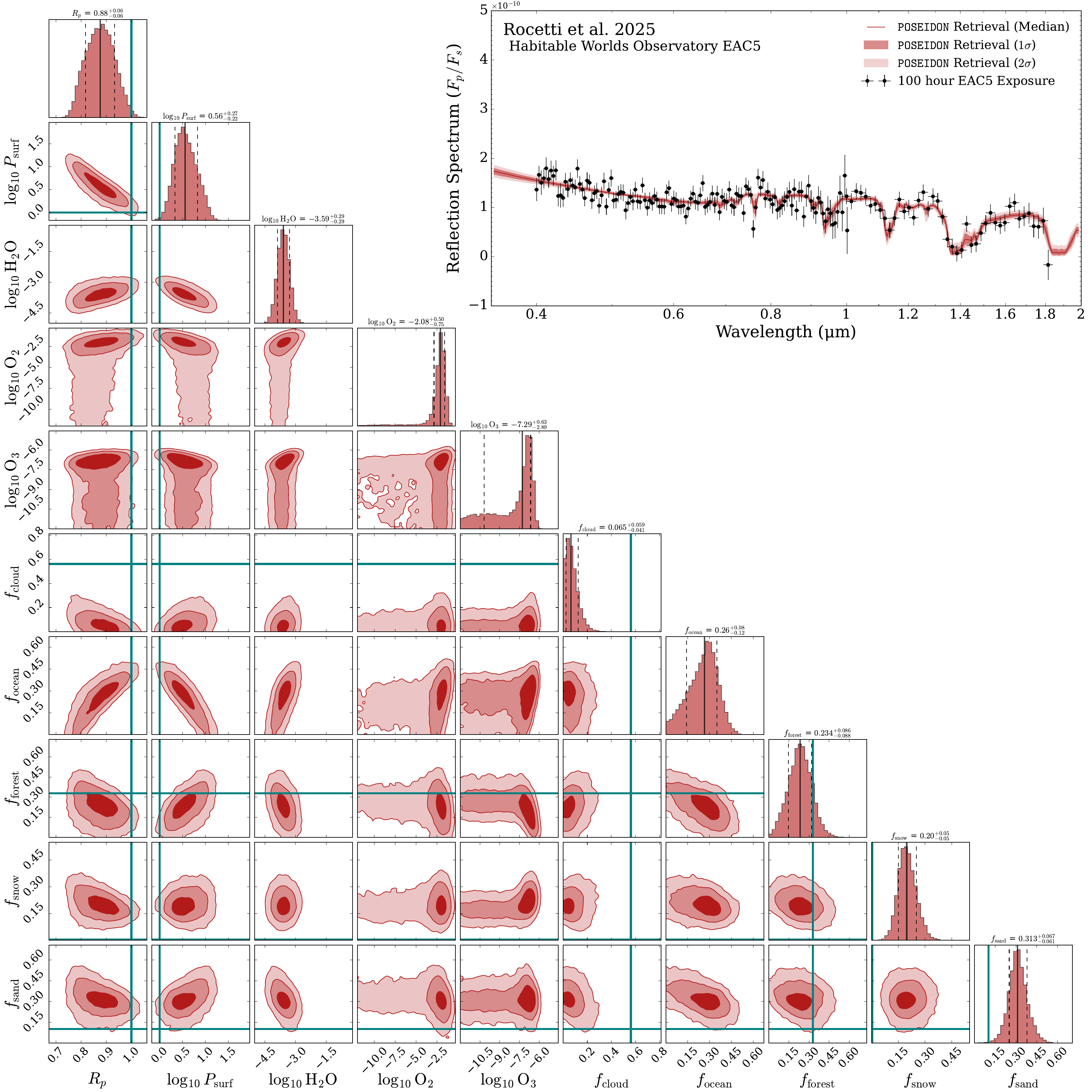}
\end{tabular}
\end{center}
\caption 
{ \label{fig:retmystic} Retrieval results for the spectral model from \citet{roccetti_planet_2025}. The posterior distributions for selected parameters are shown on the left-hand side, along with the median-retrieved spectra and simulated data in the top-right panel. The marginalized median and 1-$\sigma$ intervals for each parameter are labeled above the distributions and are shown in the black solid and dashed lines, respectively. The model parameterization provided by \citet{roccetti_planet_2025} for specific variables, discussed in Sec.~\ref{sec:mystic}, is shown by the solid teal line in the posteriors. The full posterior distribution is shown in \reffig{fig:cornerMYSTIC}.} 
\end{figure*}

Results from the \citet{roccetti_planet_2025} spectral model retrieval are shown in \reffig{fig:retmystic}.
We find that the nested sampling required $f_{\rm cloud}=0.065^{+0.059}_{-0.041}$ to reproduce the flux of the planet, despite their ERA5-based clouds covering 52.9\% of the model.
In \citet{roccetti_planet_2025}, the authors point out how zooming out their 3D spectral grid to a single pixel (1D) spectra brightens their model.
Thus, our 1D retrieval model may be systematically brighter than the 3D model used in \citet{roccetti_planet_2025}, requiring lower cloud coverage to reproduce the flux.
The marginalized median planetary radius of $R_p=0.88^{+0.06}_{-0.06}~R_{\oplus}$ from the retrieval is underestimated with respect to Earth.
Atmospheric water vapor abundance was tightly constrained to $\logt {\rm H_2 O}=-3.59^{+0.29}_{-0.29}$.
O$_2$ was constrained to $\logt {\rm O_2}=-2.08^{+0.50}_{-0.75}$ and and upper limit on O$_3$ was placed of $\logt {\rm O_3} = -7.29^{+0.63}_{-2.89}$.

For the planetary surface, a liquid water ocean was detected at $f_{\rm ocean}=0.26^{+0.08}_{-0.12}$, inconsistent with the input surface ($f_{\rm ocean}=0.56$).
Sand and snow coverage were significantly overestimated with $f_{\rm snow}=0.20^{+0.05}_{-0.05}$ and $f_{\rm sand}=0.313^{+0.067}_{-0.061}$ rather than the provided fractional coverages of 0.004 and 0.101, respectively.
We underestimated the total forest cover, with the retrieved fraction of $f_{\rm forest}=0.234^{+0.086}_{-0.088}$ just short of 1-$\sigma$ of the reference coverage of 0.329.
The lower estimate of forest coverage is likely partly due to differences in our surface-spectral choices compared with those used by {HAMSTER}~\citep{roccetti_hamster_2024}.
{Laboratory and field measurements of individual materials are often conducted under ideal conditions, and do not capture the complex structure present in real observations of planets.
\citet{roccetti_planet_2025} found that using biota spectra from USGS and ECOSTRESS spectra alone, which are common in Earth modeling and shown in \reffig{fig:ptx}, can overestimate the reflectance by up to a factor of $\sim 2$ in key bands compared to the more realistic HAMSTER models that account for this spatial complexity.}
Thus, our forest fraction likely needed to be lower to account for this difference.

\section{Discussion} \label{sec:discussion}

We demonstrated that surface composition can be weakly constrained using atmospheric retrieval algorithms for single-visit VIS/NIR HWO EAC 5 spectroscopic observations.
Previous analyses that retrieved wavelength-independent surface albedos consistently constrain surface pressure~\citep{feng2018, salvador2025, mullens2026}.
Other research using agnostic or lab-based surface prescriptions has constrained surface pressure by employing broader spectral coverage in the UV and NIR~\citep{gomez_barrientos_search_2023, ulses_detecting_2025} and by using noise fixed to the true flux values~\citep{gomez_barrientos_search_2023,wang_unveiling_2022, ulses_detecting_2025}.
For our VIS/NIR models (0.4-1.8 ${\rm \mu m}$) with Gaussian scattered noise, we find that the surface pressure is overestimated across all observation models and is degenerate with many free parameters in the retrieval, notably $R_p$ and $f_{\rm cloud}$.

Clouds play a unique role in both distinguishing and suppressing spectral signatures when modeling modern Earth's atmosphere.
While clouds significantly enhance the brightness of the planet, as can be clearly seen in {\reffig{fig:obsCloud}}, they mask surface features and can be degenerate with flatter surface spectra.
Because clouds on Earth are near the surface in pressure space, they act as an opaque reflecting surface in the HWO wavelength range.
They significantly complicate the classification of surface components on planets with moderate global cloud cover because they reflect photons that would otherwise interact with the planetary surface.
Depending on cloud properties, such as height, thickness, aerosol abundance, and molecular size, any constraints on surface properties for a planet with significant cloud coverage may be unreliable.
The planetary surfaces covered by optically thick H$_2$O clouds will not contribute to the observed spectra.
On Earth, optically thick water clouds ($\tau_{\rm H_2O}>2$) cover a global average of $f_{\rm cloud}=0.56$ of the planet~\citep{stubenrauch2013}.
Note that clouds cover the ocean around 10-15\% more than land~\citep{stubenrauch2013}, potentially causing an overestimation in land coverage for patchy observations of Earth-like planets.
However, we find the brighter continuum from thick, low-altitude clouds, which enabled tighter constraints on gas-phase absorption features above the cloud deck because gas-phase absorption generally imprints larger signals on highly reflective cloudy spectra than on cloudless ones.

The condensation of water vapor in Earth's atmosphere plays an essential role in Earth's energy balance and radiative forcing~\citep{Hartmann1992,kaltenegger_how_2017, harrison1990seasonal, ramanathan1989, wielicki1995}.
As shown in \reffig{fig:ptx}, gas-phase H$_2$O is removed from the atmosphere as it condenses into clouds.
Previous work~\citep{damiano2022} attempted to account for this by introducing a condensation ratio, which reduces the volume mixing ratio of the gaseous species condensing into the atmosphere.
Ref.~\citet{damiano2022}'s study of terrestrial planets, however, found it difficult to constrain the H$_2$O condensation ratio.
Further developing self-consistent techniques to integrate gas- and solid-phase abundances for future observations of planets in the HZ will be crucial for understanding planetary cycles, similar to Earth's water cycle.

There are significant degeneracies in the retrievals from all modern Earth-reflectance observations.
In the absence of clouds, the planetary radius shows significant degeneracies with surfaces that lack characteristic spectral features in the VIS and NIR, such as the ocean and sand.
We find that the combination of degeneracies significantly underestimates ocean coverage across all models, and often significantly overestimates sand coverage.
Additional polarimetric phase curves may enable better-informed priors on ocean coverage in retrievals by detecting ocean glint~\citep{robinson2010, vaughan_chasing_2023, cowan2025}.
The planet-star contrast ratio scales as $F_p/F_s \propto R_p^2 A_g(\lambda)$, so two scenarios are possible: either a small planet with a bright surface or a large planet with a dim surface.
That is, an underestimation of $R_p$ can be compensated by adjusting the surface composition; increasing $f_{\rm sand}$ while decreasing $f_{\rm ocean}$ can normalize the overall flux to the decrease in radius.
We find that the retrieval underestimates $R_p$ for all spectral models, except the 100\% cloud-coverage model.
The degeneracies present on $R_p$ are analogous to studies of brown dwarf spectra, which find the radius difficult to constrain within expected physical limits~\citep{burningham2021,hood2023,kitzmann2020,lueber2022,ceva2025}.

Retrievals from VIS and NIR observations of the planet alone, for a single 100-hour visit, can not yet identify land fraction coverage sufficiently to break the degeneracy between surface albedo and planetary radius without \emph{a priori} information about $R_p$.
An opaque cloud deck can also produce a similar effect: $F_p/F_s$ increases with $f_{\rm cloud}$ and is degenerate with surface coverage.
In the posterior distributions of all models, shown in Figs.~\ref{fig:corner0cloud}-\ref{fig:cornerMYSTIC}, these degeneracies can clearly be seen.
Despite the degeneracies between surface components, we find that $f_{\rm forest}$ was the most consistently detected material due to chlorophyll's spectral signatures that are distinct from those of any other surface reflection included in our model.
However, a single constrained biota-surface coverage detection is insufficient to support claims about potential life on the planet's surface.
{Multiple surface materials could have identifiable false positive red-edge-like signals, such as iron oxide~\citep{borges_detectability_2024}, cinnabar, and sulfur~\citep{parenteau2026}.}
Similar to how biosignature gases are sought for in {combination to rule out false positives}~\citep{kaltenegger_how_2017, schwieterman_exoplanet_2018}, surface biopigments need to be assessed in context~\citep{meadows2022}.

To identify specific surface coverage using retrievals currently requires \emph{a priori} information about the planet.
In addition, as shown in the retrieval using the model from \citet{roccetti_planet_2025} in Sec.~\ref{sec:retmystic}, improper assumptions about the material used to model surface reflectivity can significantly bias the retrieved spectrum.
While work has begun on creating standard sets of spectra to use for exoplanet modeling~\citep{hegde_colors_2013, hegde_surface_2015, omalley-james_biofluorescent_2018,omalley-james_expanding_2019, roccetti_hamster_2024, coelho_color_2022, coelho_purple_2024, coelho2025}, working together to expand current databases with new measurements that cover wider wavelength ranges and determine a representative set of standard spectra is a critical next step for the community.
Future work should also explore retrievals using a large, diverse set of surface spectra to determine whether current retrieval algorithms can distinguish surface signals in reflected-light observations across a wide range of surfaces, identify degeneracies, and disfavor false positives.
{This should also include further analysis into the effect spectral heterogeneity~\citep{burr2026} and surface configuration has on surface spectra retrievals.}
Other methods, including agnostic surface-retrieval methods that retrieve parametric surface albedos~\citep{wang_unveiling_2022, gomez_barrientos_search_2023, ulses_detecting_2025, burr2026}, will be significantly less biased by choices in lab data, but cannot determine constraints on surface fraction coverage.
A combination of the two methods would likely be the most robust: first performing agnostic surface retrievals to extract key features in the planet's reflection spectra, such as a red edge-like feature, and then using those features to inform the selection of lab-based spectroscopic retrieval datasets.

\section{Conclusions} \label{sec:conclusions}
Reflected light from the surfaces of terrestrial exoplanets provides a unique opportunity to constrain the coverage of surface materials and biopigments.
However, detailed studies assessing whether current HWO designs can retrieve such surface features in single-visit observations have not been available.
Using five modern Earth spectral models at quadrature and complex noise modeling for the HWO Exploratory Analytic Case 5, we simulated single-visit HWO observations and performed spectral retrievals with the open-source code {\tt POSEIDON} to assess our ability to constrain both the surface and atmospheric compositions.

Our work suggests that state-of-the-art atmospheric retrieval algorithms currently used for exoplanet observations show promise in constraining the surface composition of reflected-light HWO observations.
However, when \emph{a priori} information about the planetary radius and surface is unavailable, we found that significant degeneracies complicate proper classification of the radius, surface pressure, surface composition, and cloud cover.
We also found that bright clouds can enhance the detection of key atmospheric species but suppress surface spectral signals, complicating the detection of surface biopigments, such as the chlorophyll-induced red edge on modern Earth.

Our analysis shows that exploring degeneracies in VIS/NIR HWO observations and the biases inherent in current reflected-light retrievals will be critical priorities for mission design and data analysis.
We find that it is critical to mature these techniques and develop concrete strategies for detecting surface features and breaking degeneracies before the first light of next-generation observatories.

\begin{acknowledgments}
A.S.Z. would like to thank the support from the New York Space Grant.
We would also like to thank Ryan MacDonald and Giulia Roccetti for helpful comments. This material is based upon work supported by the National Aeronautics and Space Administration under Grant No. 80NSSC24K0139 issued through the ROSES 2022: Astrophysics Decadal Survey Precursor Science program and through a NASA Cooperative Agreement awarded to the New York Space Grant Consortium. Any opinions, findings, conclusions or recommendations expressed in this material are those of the authors and do not necessarily reflect the views of NASA.
\end{acknowledgments}





%


\appendix    

\section{Count Rates}\label{sec:counts}

Here we describe and review the components of the photon count rates that are used to compute the SNR in \refeqn{eqn:snr}.
The count rate equations below are adapted from \citet{brown_single-visit_2005, stark_maximizing_2014, garrett2017, robinson_characterizing_2016, lustig-yaeger_coronagraph_2019}, with the forms specifically following \citet{robinson_characterizing_2016}.
The count rate from the planet is defined as 
\begin{eqnarray} \label{eqn:c_p}
    c_{\rm p}&=&  q\left(\lambda\right) \tau_{\rm core}\left(\alpha, \lambda_0 \right) \mathcal{T}\left(\lambda\right)\frac{\lambda}{hc}F_{{\rm p},\lambda}\left(d\right)\,\Delta\lambda A,
\end{eqnarray}
where $\frac{\lambda}{hc}F_{{\rm p},\lambda}(d)$ is the photon flux density of the planet at a distance $d$ away, and $\Delta \lambda$ is the observing bandpass.
For ease of reading, we hereafter shorten $q(\lambda)$, $\mathcal{T}(\lambda)$, and $\tau_{\rm core}\left(\alpha, \lambda_0 \right)$ to $q$, $\mathcal{T}$, and $\tau_{\rm core}$, respectively.
Contamination from the stellar residual is given by
\begin{eqnarray} \label{eqn:c_sr}
    c_{\rm sr} = C q \tau_{\rm core}\mathcal{T}\frac{\lambda}{hc}F_{{\rm s},\lambda}(d)\,\Delta\lambda A,
\end{eqnarray}
where $C$ is the observatory's raw contrast.
This is effectively the stellar count rate modified by the observatory's contrast.

The zodiacal light within {\tt coronagraph} is computed as
\begin{eqnarray} \label{eqn:c_z_coro}
    c_{\rm z}= A q \tau_{\rm core}\mathcal{T} \Omega \Delta\lambda\,\frac{\lambda}{hc}\frac{F_{\odot,\lambda}(1\ \mathrm{AU})}{F_{\odot,V}(1\ \mathrm{AU})}\,F_{0,V}\,10^{-0.4M_{z,V}}
\end{eqnarray}
where $F_{0,V} = 3.6 \times 10^{-8} \rm{~Wm}^{-2} \mu \rm{m}^{-1}$ is the standard zero-magnitude V-band specific flux density and $F_{\odot, V}$ is the solar flux density in V-band.
Similarly, $F_{\odot, \lambda}$ is the wavelength-dependent solar flux density.
The V-band surface brightness of the zodi in units of magnitudes per unit solid angle is given by $M_{{\rm z}, V}$, where the median V-band brightness at a longitude of 135 degrees is 23 mag arcsec$^{-2}$~\citep{stark_maximizing_2014}, which we adopt in our work.
The photometric aperture area is $\Omega$ with units of arcsec$^{-2}$.
Similar to how we computed $\tau_{\rm core}$ in Sec.~\ref{sec:CI}, we assume a circular photometric aperture with radius $\sqrt{2}/2~\lambda/D$, yielding a photometric aperture area of $\Omega=\pi/2~(\lambda/D)^2$.
The exozodiacal dust count rate is similarly parameterized to the local zodiacal dust count rate, as
\begin{eqnarray} \label{eqn:c_ez}
    c_{\rm ez}= A q \tau_{\rm core} \mathcal{T} \Omega \Delta\lambda\,\frac{\lambda}{hc} \left( \frac{1 \mathrm{AU}}{r}\right)^2 \frac{F_{{\rm s},\lambda}(1\ \mathrm{AU})}{F_{{\rm s},V}(1\ \mathrm{AU})} \frac{F_{{\rm s},V}(1\ \mathrm{AU})}{F_{\odot,V}(1\ \mathrm{AU})}  \,F_{0,V}\,N_{\rm ez} 10^{-0.4M_{{\rm ez}, V}},
\end{eqnarray}
where $r$ is the exoplanet's orbital distance.
Similarly to Equation~\ref{eqn:c_z_coro}, $M_{{\rm ez},V}$ is the V-band surface brightness of the exozodical disk in units of mag arcsec$^{-2}$ where $N_{\rm ez}$ is the number of exozodis in the disk.
For our work, we assume 1 zodi ($N_{\rm ez}=1$) and an exozodiacal surface brightness of $M_{{\rm ez},V}=22$ mag arcsec$^{-2}$ for a solar twin~\citep{stark_maximizing_2014}.

From the specific detectors aboard the observatory, the count rate sourced from the dark current is defined by
\begin{eqnarray} \label{eqn:c_DC}
    c_{\rm DC} = n_{\rm pix} D_{e^-},
\end{eqnarray}
where $n_{\rm pix}$ is the number of pixels used inside the photometric aperture and $D_{e^-}$ is the dark current in electrons per pixel per unit time.
We assume the spectrum is sampled at $\Delta N_{\rm pix}=3$ pixels per lenslet, where the diffraction limit at the minimum wavelength of the observing mode $\lambda_{\rm min}$ determines the lenslet diameter, $\theta_{\rm lens} =\lambda_{\rm min}/D/2$.
Following \citet{robinson_characterizing_2016} (see {Eqn.~A17}), the number of pixels for the IFS is computed using $n_{\rm pix} = 8 \Delta N_{\rm pix} \Omega/\pi \theta_{\rm lens}^2$.

Similarly, the count rate due to the read noise (RN) is given as
\begin{eqnarray} \label{eqn:c_RN}
    c_{\rm RN} = n_{\rm pix} \frac{R_{e^-}}{ t_{\rm read}}
\end{eqnarray}
where $R_{e^-}$ is the read noise in electrons per pixel and $ t_{\rm read}$ is the time of each readout.
Count rates from the clock-induced charge (CIC) are given by
\begin{eqnarray} \label{eqn:c_CIC}
     c_{\rm CIC} = n_{\rm pix} \frac{R_{\rm c}}{ t_{\rm read}},
\end{eqnarray}
where $R_{\rm c}$ is the detector-specific CIC in electrons per pixel and assume $t_{\rm read}=1000$ s~\citep{the_luvoir_team_luvoir_2019} for both the CIC and RN.
Thermal noise contributions from the observatory are also included,
\begin{eqnarray} \label{eqn:c_th}
    c_{\rm th} =  q \Omega \tau_{\rm core}\mathcal{T} \frac{\lambda}{h c} \epsilon_{\rm sys} B_{\lambda}(T_{{\rm sys}})\,\Delta\lambda A,
\end{eqnarray}
where $B_{\lambda}(T_{{\rm sys}})$ is the Planck blackbody function for a telescope of temperature $T_{\rm sys}$ and effective emissivity $\epsilon_{\rm sys}$, which is of order unity.
We assume the observatory is kept at an ambient temperature of $T_{\rm sys}=270$ K.

We add a term into the SNR equation similar to \citet{garrett2017} that accounts for variation in the residual stellar speckle
\begin{eqnarray} \label{eqn:c_sp}
    c_{\rm sp} = c_{\rm sr} f_{\rm pp},
\end{eqnarray}
where $f_{\rm pp}$ is a post-processing factor.
The post-processing factor determines the stability of the PSF, such that $f_{\rm pp}=0.1$ corresponds to a 10-fold reduction in the stellar residual, which is the value we assume for our analysis.
We note how $c_{\rm sp}$ is not scaled by the exposure time in \refeqn{eqn:snr}, setting the noise floor of our observations in the limit 
\begin{eqnarray} \label{eqn:snrlim}
\lim_{t_{\rm exp}\to\infty} {\rm SNR}=c_{\rm p}/c_{\rm sp}.
\end{eqnarray}
Thus, even for infinitely long exposures, we find that resolving the planetary signal remains dependent on the ability to reduce and stabilize the speckle residual PSF.

\begin{figure}[htbp]
\begin{center}
\begin{tabular}{c}
\includegraphics[width=\linewidth]{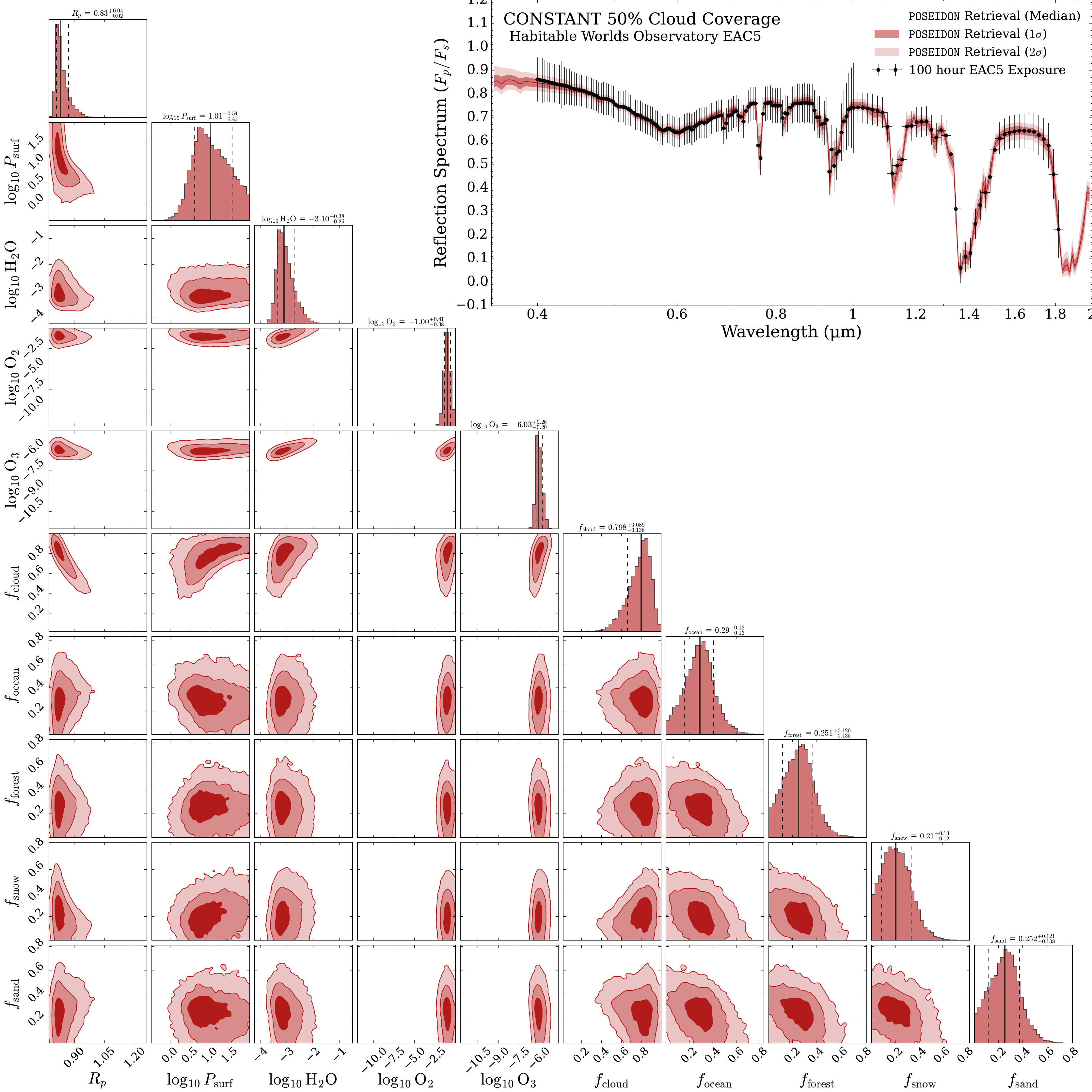}
\end{tabular}
\end{center}
\caption 
{ \label{fig:CONST50Ret} Retrieval results for the self-consistent {\tt POSEIDON} spectra with 50\% cloud coverage with fixed data on the true flux values. The posterior distributions for selected parameters are shown on the left-hand side, along with the median-retrieved spectra and simulated data in the top-right panel. The marginalized median and 1-$\sigma$ intervals for each parameter are labeled above the distributions and are shown in the black solid and dashed lines, respectively.} 
\end{figure}

{
\section{Fixed Noise Retrieval} \label{sec:fixed}

To ensure single noise draws from simulating observations with Gaussian noise are not driving the retrieval results, it is often recommended to perform a suite of retrievals on numerous noise realizations.
However, reflected light retrievals can take thousands of hours of compute time each, making this analysis unfeasible.
As \citet{feng2018} noted, a compromise is to fix the data on the true flux values and add error bars to the data points scaled by the SNR.
We performed a retrieval analysis of the $f_{\rm cloud}=0.5$ model described in Sec.~\ref{sec:poseidonspec} while fixing the data points to the true flux values.
The results of this retrieval are shown in \reffig{fig:CONST50Ret}.
We find the degeneracies present in this fixed noise model are consistent with those discussed in Sec.~\ref{sec:ret50cloud}.
There is a significant degeneracy between $R_p$ and $f_{\rm cloud}$, where the reported radius is significantly underestimated with an overestimation of cloud coverage.
}

\begin{figure}[htbp]
\begin{center}
\begin{tabular}{c}
\includegraphics[width=\linewidth]{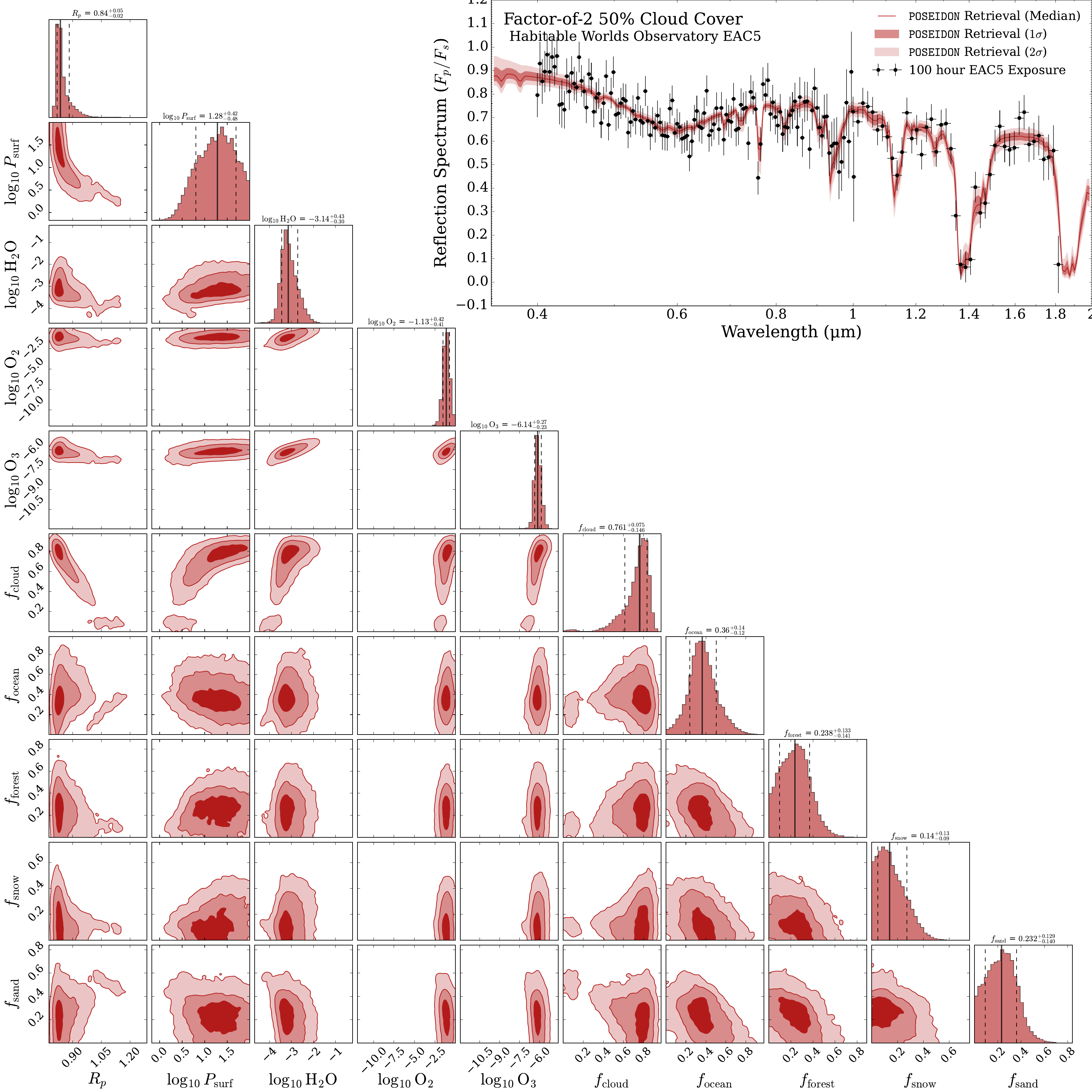}
\end{tabular}
\end{center}
\caption 
{ \label{fig:limR50Ret} Retrieval results for the self-consistent {\tt POSEIDON} spectra with 50\% cloud coverage and tighter radius prior of $\mathcal{U}(0.5,2.0)$. The posterior distributions for selected parameters are shown on the left-hand side, along with the median-retrieved spectra and simulated data in the top-right panel. The marginalized median and 1-$\sigma$ intervals for each parameter are labeled above the distributions and are shown in the black solid and dashed lines, respectively.} 
\end{figure}

{
\section{Factor-of-2 Radius Prior Retrieval} \label{sec:limR}

We ran a retrieval with a tighter uniform prior on $R_p$ of $\mathcal{U}(0.5,2.0)$ to assess where the constrained parameter space influences the retrieval results.
We performed this analysis using the $f_{\rm cloud}=0.5$, Gaussian-scattered spectral model described in Sec.~\ref{sec:poseidonspec}.
The retrieval results for the factor-of-2 radius prior are shown in \reffig{fig:limR50Ret}.
The tighter uniform radius prior does not provide any significant improvement in the performance of the retrieval.
We find the shape of the posterior distribution of $R_p$ is similar to the retrievals in Sec.~\ref{sec:ret50cloud} which uses the conservative $\mathcal{U}(0.1,10)$ radius prior.
This is likely due to the fact that the posterior distribution of $R_p$ in \reffig{fig:ret50cloud} is fully constrained within the range of $0.5$ to $2.0$ $R_\oplus$.
Thus, further \textit{a priori} information on the radius distribution is needed to provide ample constraints.
Future work should investigate performing retrievals coupled with photometric observations of the planet at different points in its orbit to determine if non-uniform constraints on the radius can be placed.
This would also require an analysis of the potential biases this introduces into constraining other parameters in the retrieval algorithm, which is outside the scope of this work.

}

\section{Retrieval Corner Plots}

\begin{figure}[htbp]
\begin{center}
\begin{tabular}{c}
\includegraphics[width=\linewidth]{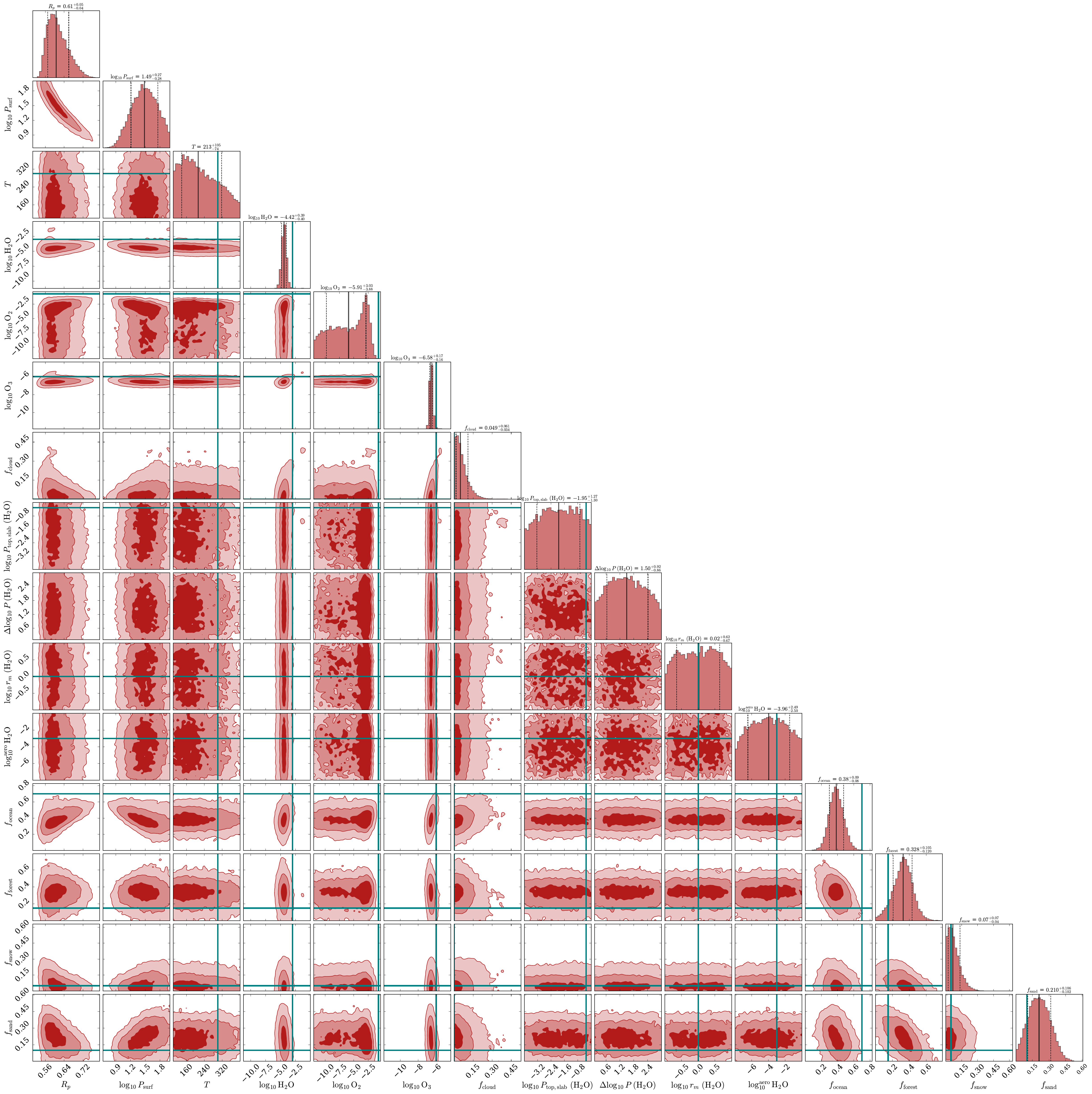}
\end{tabular}
\end{center}
\caption 
{ \label{fig:corner0cloud} Full corner plot of the 0\% cloud coverage, self-consistent {\tt POSEIDON} model retrieval.} 
\end{figure}

\begin{figure}[htbp]
\begin{center}
\begin{tabular}{c}
\includegraphics[width=\linewidth]{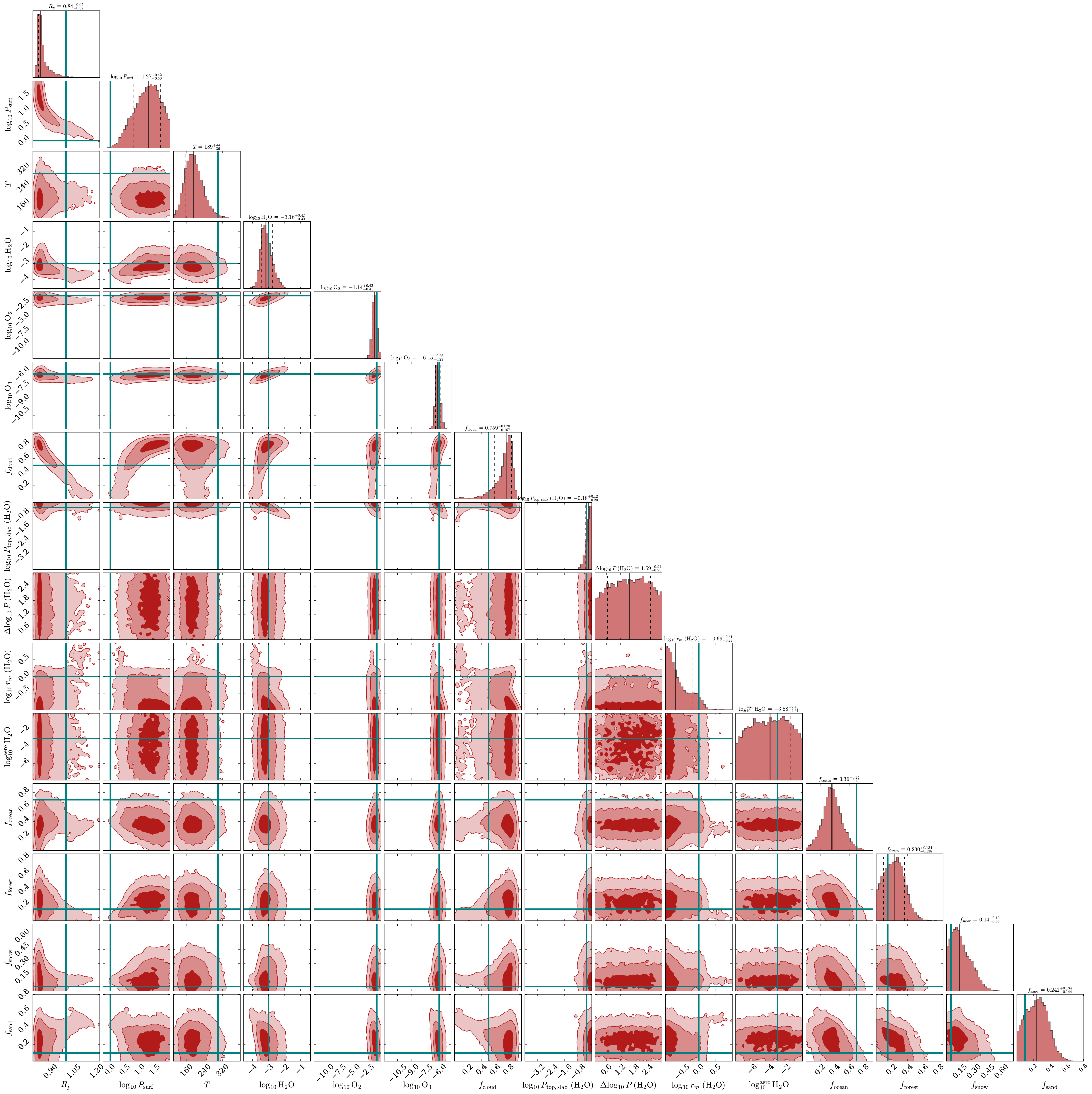}
\end{tabular}
\end{center}
\caption 
{ \label{fig:corner50cloud} Full corner plot of the 50\% cloud coverage, self-consistent {\tt POSEIDON} model retrieval.} 
\end{figure}

\begin{figure}[htbp]
\begin{center}
\begin{tabular}{c}
\includegraphics[width=\linewidth]{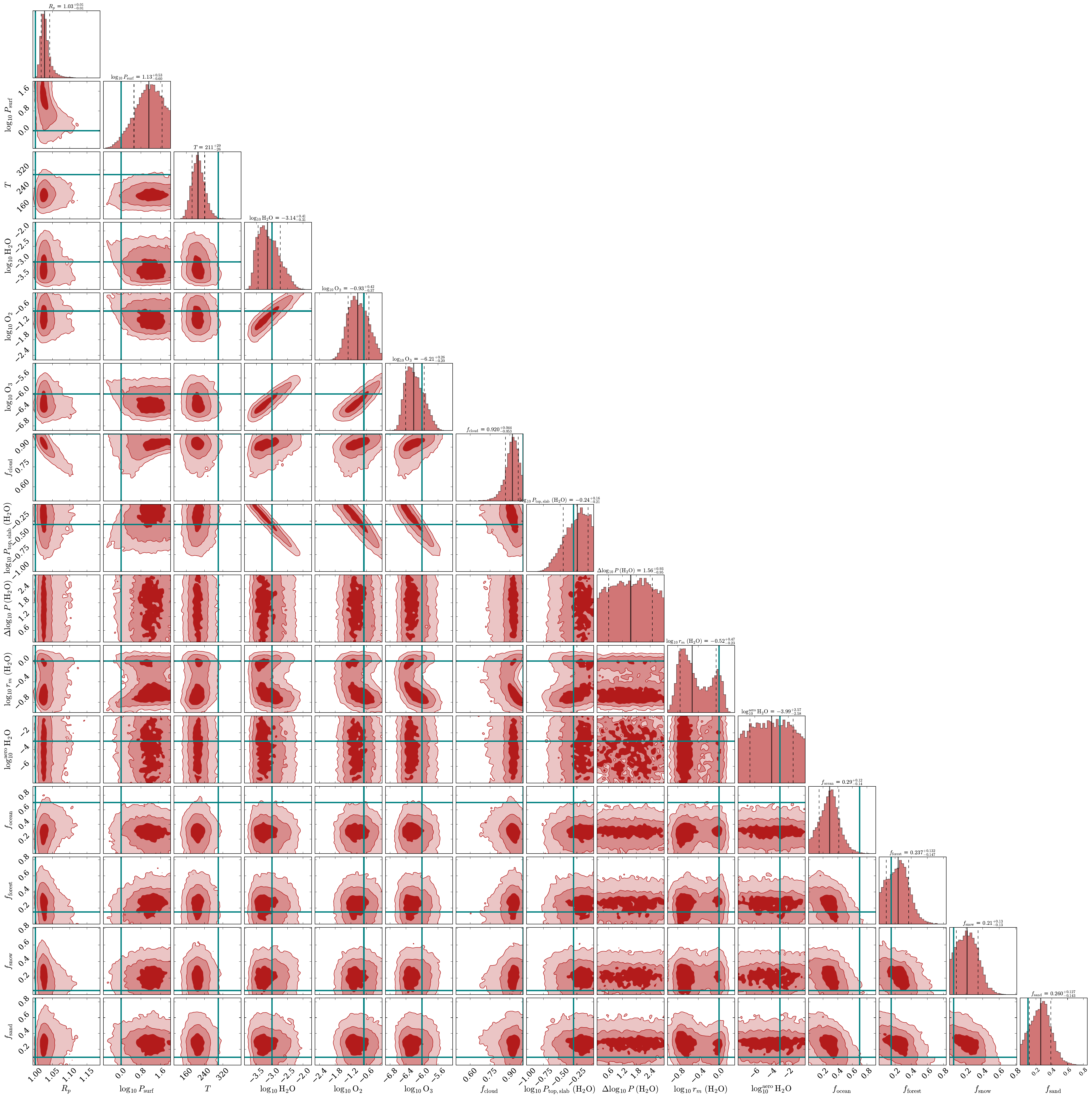}
\end{tabular}
\end{center}
\caption 
{ \label{fig:corner100cloud} Full corner plot of the 100\% cloud coverage, self-consistent {\tt POSEIDON} model retrieval.} 
\end{figure}

\begin{figure}[htbp]
\begin{center}
\begin{tabular}{c}
\includegraphics[width=\linewidth]{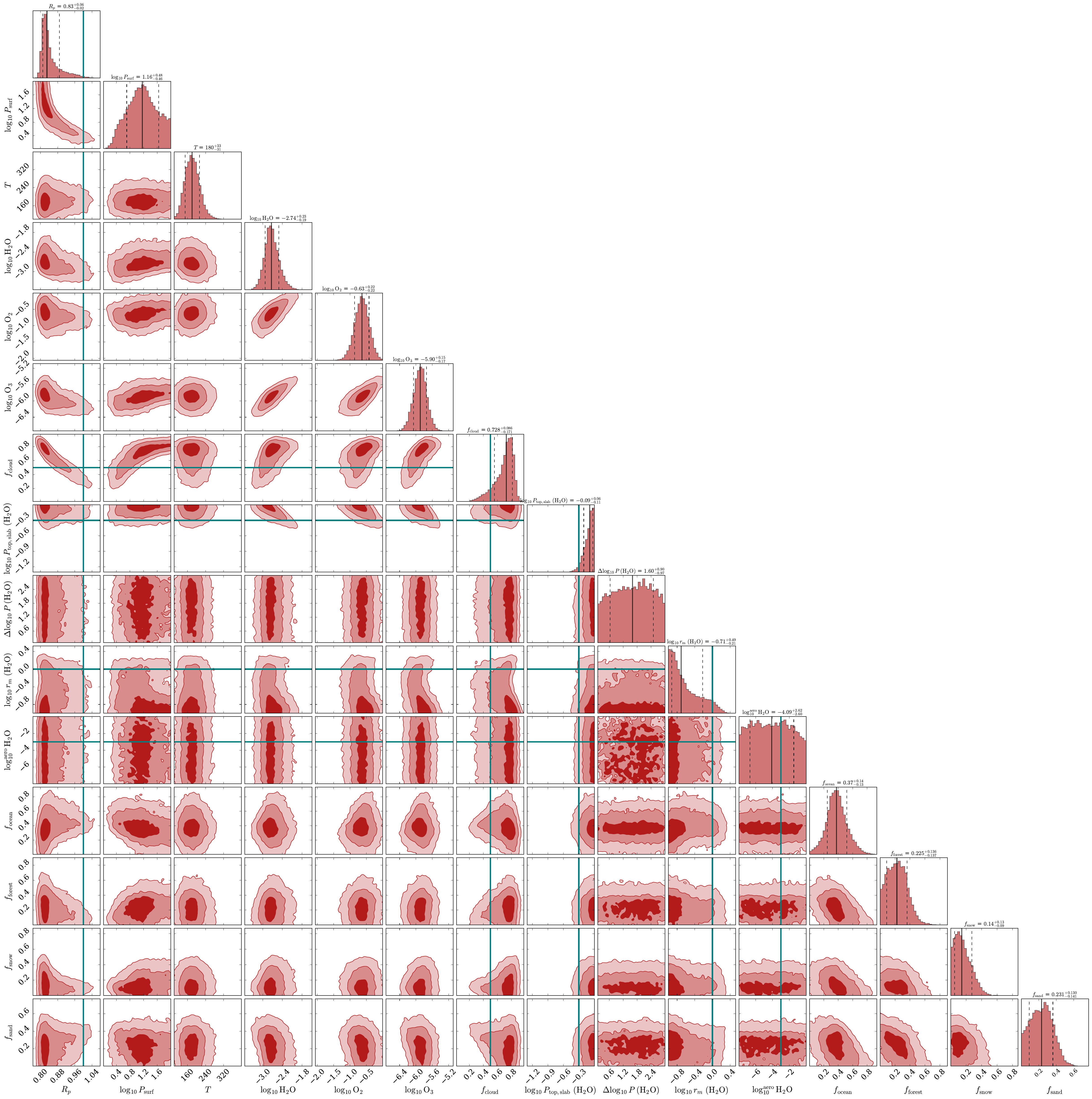}
\end{tabular}
\end{center}
\caption 
{ \label{fig:cornerMODIS} Full corner plot of the MODIS surface and \citet{kaltenegger_high-resolution_2020} atmospheric model retrieval.} 
\end{figure}

\begin{figure}[htbp]
\begin{center}
\begin{tabular}{c}
\includegraphics[width=\linewidth]{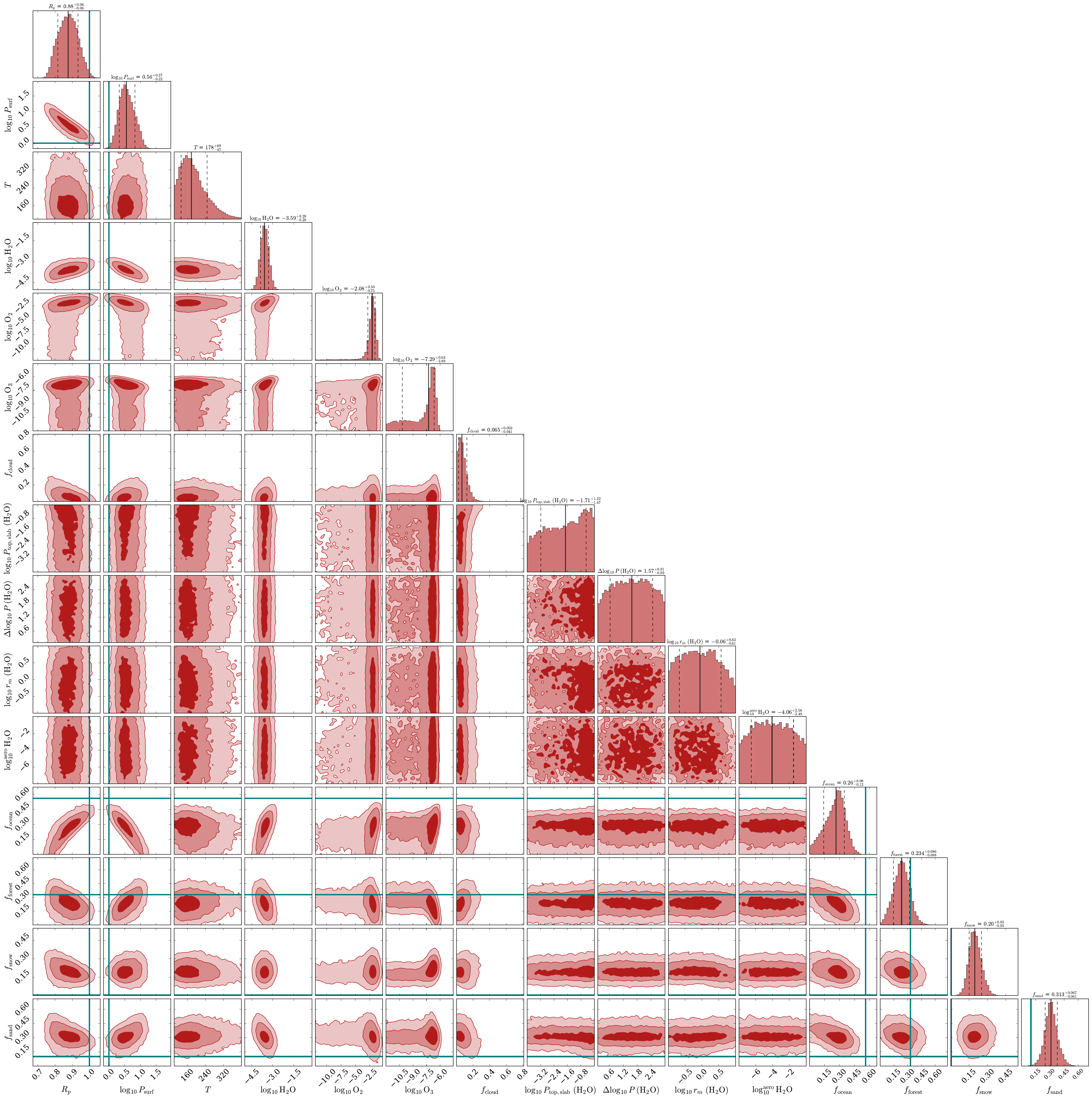}
\end{tabular}
\end{center}
\caption 
{ \label{fig:cornerMYSTIC} Full corner plot of the \citet{roccetti_planet_2025} model retrieval.} 
\end{figure}

Figs.~\ref{fig:corner0cloud}-\ref{fig:cornerMYSTIC} show the full corner plots for the retrievals.
The posterior median and 1-$\sigma$ limits are shown above the marginalized distribution for each parameter.
Where possible, the reference values used to create the spectra are included in teal lines in the posteriors.


\bibliography{references, report}{}
\bibliographystyle{aastex701}



\end{document}